\documentclass[aps,twocolumn,pra,nofootinbib,preprintnumbers,amsmath,amssymb,superscriptaddress]{revtex4}

\usepackage{graphicx}
\usepackage{epsfig,amsfonts,amssymb}
\usepackage{color}
\pdfoutput=1
\usepackage{hyperref}
\usepackage{amsmath,amssymb,bm,bbm,amsfonts}

\usepackage{graphicx}
\usepackage{yfonts}
\usepackage{enumerate}
\usepackage{color}

\newcommand{\bea}{\begin{eqnarray}}
\newcommand{\eea}{\end{eqnarray}}

\newcommand{\Tr}{\mathop{\mathrm{Tr}}}

\def\Tr{\hbox{Tr}}
\newcommand{\be}{\begin{equation}}
\newcommand{\ee}{\end{equation}}
\newcommand{\beas}{\begin{eqnarray*}}
\newcommand{\eeas}{\end{eqnarray*}}

\def\Biggg#1{{\hbox{$\left#1\vbox to 25pt{}\right.\n@space$}}}
\def\n@space{\nulldelimiterspace=0pt \m@th}
\def\m@th{\mathsurround = 0pt}

\newcommand{\CR}{\nonumber \\}

\begin{document}

\preprint{YITP-13-49}

\title{Brick Walls for Black Holes in AdS/CFT}

\author{Norihiro Iizuka}
\email[]{iizuka@yukawa.kyoto-u.ac.jp, iizuka@phys.sci.osaka-u.ac.jp}  
\thanks{\\Address after April 1, 2014: {\itshape\footnotesize{Department of Physics, \\Osaka University, Toyonaka, Osaka 560-0043, JAPAN}}}
\affiliation{Yukawa Institute for Theoretical Physics, 
Kyoto University, Kyoto 606-8502, JAPAN}

%
%
%
%

\author{Seiji Terashima}
\email[]{terasima@yukawa.kyoto-u.ac.jp}
\affiliation{Yukawa Institute for Theoretical Physics, 
Kyoto University, Kyoto 606-8502, JAPAN}

\begin{abstract}
We study the 't Hooft's brick wall model for black holes in a holographic context.   
The brick wall model suggests that without an appropriate near horizon IR cut-off, 
the free energy of the probe fields 
show the divergence due to the large degenerate states near the horizons.   
After studying the universal nature of the divergence in various 
holographic setting 
in various dimensions, 
we interpret the nature of the divergence in a holographic context.   
The free energy divergence is due to the large degeneracy and continuity of the 
low energy spectrum in the boundary theory at the deconfinement phase. 
These divergence and continuity should be removed by finite $N$ effects, which make the 
spectrum discrete even at the deconfinement phase. 
On the other hand, in the bulk, these degenerate states are 
localized near the horizon, and the universal divergence of these degenerate states   
implies that the naive counting of the degrees of freedom in bulk should be 
modified once we take into account the non-perturbative quantum gravity effects near the horizon. 
Depending on the microscopic degrees of freedom, the position, 
where the effective field theory description to count the states breaks down,  
has different Planck scale dependence. 
It also implies the difficulty to have an electron like gauge-singlet 
elementary field 
in the boundary theory Lagrangian.   These singlet fields are 
at most composite fields, because they show divergent free energy, suggesting a 
positive power of $N$ at the deconfinement phase. 

\end{abstract}


\maketitle

\tableofcontents

\noindent

\section{Introduction}

Understanding the 
quantum nature of gravity has been one of the most exciting topics   
in high energy physics, and 
black holes are touchstones of our understanding of the quantum nature of gravity.  
The recent developments of the AdS/CFT correspondence, or more broadly, 
the gauge/gravity duality \cite{Maldacena:1997re, Gubser:1998bc, Witten:1998qj}, give many new important insights about black holes. 
Recent developments of various aspects of the quantum nature of black holes, 
including their microscopic entropy counting \cite{Strominger:1996sh}, the Hawking-Page transition \cite{Witten:1998zw}, the quantum nature for information 
paradox including unitarity \cite{Maldacena:2001kr},  
are related to deepening our understanding of gauge/gravity duality. 
Moreover 
gauge/gravity duality defines quantum gravity non-perturbatively. 

The brick wall model 
is the model for black holes which 't Hooft proposed \cite{'tHooft:1984re}.  
He pointed out that the degrees of freedom near the black hole event horizon, evaluated through the 
probe field free energy or entropy, always diverge. This divergence is due to the 
infinite warped factor of the metric near the horizon.  
By requiring that diverging free energy/entropy to be finite, 
we {\it have to} introduce the near horizon effective cut-off.  
The importance of brick wall model is the necessity of this cut-off. 
Furthermore, 't Hooft pointed out that by requiring it to be the  
same order of the background black hole's one, 
the cut-off scale ends up to be Planck scale measured 
by invariant distances. 
Given the non-perturbative quantum gravity from dual field theory, 
it is very natural to ask what the brick wall model implies in the gauge/gravity setting. 
The purpose of this paper is to understand 
this point.

In this paper we revisit these brick wall model results from the dual field theory viewpoint. 
First we study the universal nature of the brick wall model in various exotic black brane 
backgrounds for the probe fields. 
These analysis are done for the probe fields not only the scalar but also fermions, which is 
either charged or not, on the background which has non-trivial IR dynamics. 
Then, we interpret the Planck-scale cut-off dependence of the brick wall models 
as the large $N$ dependence of the free energy of the probe fields. 
These analyses suggest us that in dual field theory, it is difficult to have a gauge-free singlet 
object unless it is a composite object.

In order to simplify the bulk argument, in this paper we take the limit 
where $g_s$ is fixed finite and $N$ to be very large, 
so $\lambda \equiv g_s N$ is also very 
large\footnote{This limit has been considered, see for examples \cite{Azeyanagi:2012xj, Azeyanagi:2013fla}.}. 
Note that this is different from 
't Hooft limit ($g_s \to 0$, $N\to \infty$ with $\lambda \equiv g_s N$ fixed finite). 
Therefore, in our bulk analysis 
we take both string scale $l_s$ and Planck scale $l_p$ are the same order, $l_s \sim l_p$ and 
they are very small $l_s \sim l_p \ll 1$,  where 
other macroscopic quantities including AdS scale are assumed
to be ${\cal O}(1)$. This corresponds to, in boundary, the stronger coupling limit than the 't Hooft limit since 
$\lambda/N \sim g_{YM}^2$ fixed finite with $\lambda$ very large as $\lambda \sim N$. 
Due to $g_s \sim 1$, 
in bulk very short scale distance such as string scale and Planck scale, 
it is different from usual perturbative string theory where $g_s \to 0$.  
However since we consider the limit $l_s \sim l_p \ll 1$, we have low energy effective theory description. 
In this limit, we will discuss when and where the classical gravity description breaks down. In our limit, 
both stringy effects and quantum gravity effects appear at the same level.

{Another important assumption is that 
the whole gauge theory system becomes a thermal equilibrium. 
Thermal equilibrium for the system is assumed for the gravity
side, which is also the basic assumption of the brick wall model.
This implies that we will consider only the large black hole in the AdS
space-time, where the curvature of the black hole is AdS scale.}

The organization of this paper is followings; In \S II, we briefly review the known results about 
the original brick wall model for black holes by 't Hooft and see
the probe field free energy show divergence due to the near horizon regime. 
In \S III, we study the universal nature of the brick wall in various exotic black branes 
including the recently studied 
Lifshitz black branes \cite{Taylor:2008tg, Goldstein:2009cv, Goldstein:2010aw},  
hyperscaling violating black branes \cite{Charmousis:2010zz, Iizuka:2011hg}. 
We will see its 
universal nature that under the mild conditions, probe field free energy always diverge and 
its equality with the background exotic black brane yield the Planck scale.  
Part of the results of \S III is already known in the literature \cite{Mann:1990fk}.  

In \S IV, we interpret these IR divergences from the dual field theory viewpoint. 
We interpret this universal divergence of the probe field without brick wall as an indication of difficulty to introduce the gauge free (singlet) field in the dual boundary theories. 
This is the same statement that the $SU(N)$ gauge singlet fields, such as a 
fundamental electron field, are difficult to be introduced in holographic setting, 
unless 
they are the composite fields (like meson), which implies that 
they cannot be the fundamental fields in the Lagrangian level.  
There composite fermions show divergent free energy in the large $N$ limit, which is reflected to the 
universal divergent free energy of the probe field near the horizon 
without the brick wall cut-off. The brick wall introduces the 
Planck scale dependent cut-off, and this is the same as keeping the $N$ finite. 
This is one of the main result of this paper.  

We also discuss Euclidean path integral measure and its apparent non-diffeomorphism 
invariance. However this result is already known in the literature \cite{Barbon:1994ej, Giddings:1993vj, Emparan:1994qa, de Alwis:1995cr}. 

Before we proceed, we comment on the connection on the brick wall to the entanglement entropy and also  
various related references.  
In \cite{Callan:1994py,Kabat:1994vj}, the entanglement entropy is shown to be equivalent to the 
thermal entropy in Euclid Rindler Hamiltonian, and 
based on this, the black hole entropy is interpreted as entanglement entropy associated 
with the 't Hooft's brick wall model in \cite{Callan:1994py,Kabat:1994vj}. 
In this paper, however, we take a different interpretation. 
Based on the modern viewpoint, we assume that black hole entropy has dual microscopic origin from holographic viewpoint \cite{Strominger:1996sh}. In this paper, we mainly consider the 
probe field added on top of above black holes and discuss mainly 
the implication of the probe field free energy divergence and its connection to 
dual field theory degrees of freedom.  

\section{The brick wall model by 't Hooft}
\subsection{Quantization for probe fields}

In this section, we review the original 't Hooft's brick wall model \cite{'tHooft:1984re}.
First, we review the quantization of the probe scalar fields, and then 
we evaluate the partition function and free energy. Here we consider probe scalar fields, however, 
the following discussion will be applied 
to the other fields, for example the gravitons around the background metric, 
without essential modifications.

The situation we consider is space-time where we assume the homogeneity and isotropy in fixed $r$ coordinate, where $r$ is the radial coordinate. Under such mild assumption, 
we will conduct the analysis in the generic setting.
Given these assumption, the metric is specified by $g_{tt}(r)$, $g_{rr}(r)$, $g_{\vec{x}\vec{x}}(r)$ only. 
$\vec{x}$ represents the $d$ dimensional spatial dimensions in $D= d+2 $ space-time. 
By using the $r$ coordinate redefinition, we take the gauge that $- g_{tt}(r) = (g_{rr}(r))^{-1}$, so the most generic metric in this situation is always written as  
\bea
\label{thefirstmetricansatz}
ds^2 = g_{tt}(r) dt^2 + g_{rr}(r) dr^2 + g_{\vec{x}\vec{x}}(r) d\vec{x}^2 \,.
\eea
Since we consider the black hole (black brane) geometry, we assume 
$g_{tt}(r) = 0 $ at $r = r_h$, where $r_h$ is the radial position of the horizon. 

Consider, for simplicity, the minimally coupled scalar field, with mass $m$ in these background. 
The scalar field wave equation is 
\bea
\label{wavewaveeq}
\left( \Box + m^2 \right) \phi = 0 \,.
\eea
We will consider the quantization of these fields by imposing two boundary conditions. 
Since we solve the 
second order differential equations, we have two independent solutions. One boundary condition forces us 
to take the appropriate linear combinations of the two solutions. Imposing the other boundary conditions, we obtain the quantization condition for the spectrum. Let us see this more concretely. 

First, we impose the ``UV cut-off'' for the scalar field 
\bea
\label{normalizablebc1}
\quad \phi \to 0 \,. \quad (r = L \to \infty) 
\eea
If we impose this eq.~(\ref{normalizablebc1}) for the asymptotically Anti-de Sitter (AdS) setting, 
this is nothing but the requirement that the scalar field $\phi$ has no non-normalizable mode 
in the asymptotic AdS region for the positive mass scalar, $m \ge 0$ case. 
Remember that asymptotic AdS$_D$, the scalar field behaves 
\bea
\quad \quad \phi \sim r^{\Delta_{\pm}} \,, \quad \Delta_{\pm} = D/2 \pm \sqrt{(D/2)^2 + m^2} \,.
\eea 
Instead, if we impose eq.~(\ref{normalizablebc1}) in the asymptotic flat case, 
this is the same as imposing the Dirichlet boundary condition at the spatial infinity, or 
normalizable condition for the fields.  

Second, we impose the ``IR cut-off'' for the scalar field as 
't Hooft \cite{'tHooft:1984re},
\bea
\label{'tHooft'soriginalDirichlet1}
\quad \phi = 0 \,, \quad (r = r_h + h)
\eea
where $r = r_h$ is the horizon of the black hole and 
$h > 0$ is some very small distance scale which we will determine later. 
This imposes the Dirichlet boundary condition for the scalar field near the horizon.  
Actually we will see later  
that the Dirichlet boundary condition is not the crucial, we can impose 
either Dirichlet boundary condition with any constant values $C_{r_h + h}$ 
\bea
\quad \phi = C_{r_h+h} \,, \quad (r = r_h + h)
\eea
or instead Neumann boundary condition at $r = r_h + h$. 
For a moment, let's first consider the 
Dirichlet boundary condition, eq.~(\ref{'tHooft'soriginalDirichlet1}). 

Given the two boundary conditions (\ref{normalizablebc1}) and (\ref{'tHooft'soriginalDirichlet1}) 
for the second order differential equation, 
the allowed mode is always quantized, and its spectrum $E$ is parameterized by 
discrete parameters $(m,\vec{k})$ and integer $n$.  

Let us see this more concretely; 
Taking the ansatz, 
\bea
\label{phiansatzveck}
\phi = \phi(r) e^{- i E t  + i \vec{k} \vec{x} }
\eea
the wave equations for $\phi(r)$, 
\bea
\label{scalarwaveeq1}
\frac{1}{\sqrt{-g}} \partial_r ( \sqrt{-g} g^{rr} \partial_r \phi) - E^2 g^{tt} \phi - \vec{k}^2 g^{\vec{x}\vec{x}} \phi  - 
m^2 \phi = 0 \,,
\eea
allows generic solutions 
\bea
\phi(r,E, \vec{k}, m)&=& c_{1}(E, \vec{k}, m) F^{(1)}(r,E, \vec{k}, m) 
\CR &&+  c_{2}(E, \vec{k}, m) F^{(2)}(r,E, \vec{k}, m)  \,,
\eea
where $F^{(1)}(r,E, \vec{k}, m) $ and $F^{(2)}(r,E, \vec{k}, m)$ are two independent solutions 
for the equation (\ref{scalarwaveeq1}) and 
$c_{1}(E, \vec{k}, m)$ and $c_{2}(E, \vec{k}, m)$ are constant w.r.t. $r$ coordinate. 

We have freedom to re-define any linear combinations of $F^{(1)}$ and $F^{(2)}$ as new $F^{(1)}$ and $F^{(2)}$, so by using this freedom, we can always take the choice such that $F^{(1)}$ 
satisfy the UV boundary condition eq.~(\ref{normalizablebc1}). 
Generically for this choice of $F^{(1)}$, $F^{(2)}$ do not satisfy the boundary condition eq.~(\ref{normalizablebc1}), therefore it forces us to set 
\bea
c_{2}(E, \vec{k}, m)  = 0 \,.
\eea
Given this, the 't Hooft's IR boundary condition (\ref{'tHooft'soriginalDirichlet1}) gives the 
quantization condition for $E$, by  
\bea
\label{quantizationcondition1}
F^{(1)}(r = r_h + h,E, \vec{k}, m)  = 0 \,. 
\eea
This condition yields discreteness for the energy eigenvalues $E$. 
Here we have assumed that we have also the IR cut-off along the $\vec{x}$ directions therefore, 
$\vec{k}$ is also quantized. 
We label the discrete energy eigenvalues satisfying condition eq.~(\ref{quantizationcondition1}) as 
$E_n$, where $n$ is positive integer and we take $E_n < E_n'$ for $n < n'$.

\subsection{Partition function for probe fields}

Given the discrete spectrum for the scalar field around black hole background, we will consider the 
canonical ensemble for this scalar field with temperature given by the Hawking temperature of the 
black hole $1/\beta$. 

For that purpose, it is convenient to introduce the occupation number $g(E)$ below the 
energy $E$.  
First, we take the ansatz 
\bea
\phi = \phi_0 e^{- i E t + i \int k_r(r) dr +i \vec{k} \vec{x} } \,,
\eea
where $\phi_0$ is constant, 
which is the same as setting 
\bea
\label{krphi0relation}
\phi(r) = \phi_0 e^{ i \int k_r(r) dr}  
\eea
to define $k_r(r)$ for $\phi(r)$ in eq.~(\ref{phiansatzveck}). Then 
we have quantization conditions from eq.~(\ref{quantizationcondition1})
\bea
\label{quantizationnpi}
n \pi  = \int_{r_h + h}^{L}  k_r(r, E_n, \vec{k}) dr \quad \mbox{(for positive integer $n$)} \,, \quad
\eea
with $L \to \infty$. 
This gives the discrete spectrum labeled by $n$, $\vec{k}$, $h$, $m$ as 
\bea
\label{discreteElabeledbyn}
E = E_n (\vec{k}, h, m)  \,.
\eea

Therefore, by integrating over the $\vec{k}$ modes, the occupation number $g(E)$ is given by 
\bea
\label{gEdefinition}
g(E) = \frac{V}{\pi } \int  k_r(r, \vec{k} , E) dr d\vec{k}
\eea
where $V$ is proportional to the volume of the field theory spatial dimensions 
defined by 
\bea
V \equiv 
\frac{1}{(2 \pi)^d} \int d \vec{x} \,, 
\eea
and it excludes $r$ direction\footnote{In holographic setting $r$ direction does not appear in the 
dual field theory.}. Here we have also approximated that the volume is large enough so that 
the mode summation is given by $\sim V d \vec{k}$.  

Then we find that the  
partition function for the scalar field $\phi$ is given by 
\bea
Z_{\phi} &=&  
\prod_{\vec{k}}
\prod_{n} \sum_{m=0}^\infty e^{- \beta m E_n} 
\CR &=&  
\prod_{\vec{k}}
\prod_{n} \frac{1}{1 - e^{- \beta  E_n}} \,,
\eea
where 
$n$ is for discrete spectrum, 
$\vec{k}$ in $\prod_{\vec{k}}$ is taken for the integer quantum numbers  
and the second 
summation over $m$ is due to the boson. Therefore, 
\bea
\beta F_{\phi} &=& V \, \int d\vec{k}  \sum_n \, \log \left( 1 - e^{- \beta E_n} \right) \, \nonumber \\
&=& \int dE  \left(\frac{\partial g(E)}{\partial E}\right) \log \left( 1 - e^{- \beta E} \right) \,, \nonumber \\
 &=&- \int dE \frac{\beta g(E) }{\left( 1 - e^{- \beta E} \right)} \,, \nonumber \\
 &=&- \frac{\beta V}{\pi} \int \frac{dE}{\left( 1 - e^{- \beta E} \right)} dr \left( d\vec{k}  k_r(r, \vec{k} , E) \right) \,, \quad
\label{betafreeenergy}
\eea
where $\vec{k}$ is approximate as the continuous variables.
Given the wave equations for $k_r(r, \vec{k} , E)$, we can evaluate this. 
There are trivial volume dependence (\ref{betafreeenergy}) in above expression, which we will 
omit since we will always consider free energy/entropy per unit volume. 

We have considered the specific Dirichlet boundary condition for 
both UV and IR boundary condition in eq.~(\ref{normalizablebc1}) and 
(\ref{'tHooft'soriginalDirichlet1}). 
For the free energy evaluation, the effects of the boundary condition only affect the 
quantization condition eq.~(\ref{quantizationnpi}), (\ref{discreteElabeledbyn}), (\ref{gEdefinition}). If we consider more generic Dirichlet boundary condition, 
where in UV, we choose $\phi = C_{\infty}$, and in IR, we choose $\phi = C_{r_h + h}$, then, 
quantization condition is shifted as 
\bea
\quad \quad {C_{\infty}}&=& e^{i \int^L_{r_h+h} dr k_r(r) + 2 \pi i n}\, {C_{r_h + h}}  \CR 
\Rightarrow \quad -i \log (\frac{C_{\infty}}{C_{r_h + h}} ) 
&=& \int^L_{r_h+h}  dr k_r(r) + 2 \pi n \,.
\eea
Eq.~(\ref{quantizationnpi}) corresponds to  ${C_{\infty}}/{C_{r_h + h}}
= e^{i \pi n}$ with 
${C_{r_h + h}} \to 0$. However it is clear that these modification of the boundary conditions 
will not qualitatively change the $g(E)$ for eq.~(\ref{gEdefinition}).  
Furthermore, if we replace both UV and IR boundary condition by Neumann boundary condition, 
$g(E)$ is not modified qualitatively in that case too. In this way, for the free energy, 
these modifications of the boundary condition will not affect the argument below qualitatively.

Using the WKB approximation, from eq.~(\ref{scalarwaveeq1}) and (\ref{krphi0relation}), we obtain 
\bea
\label{krdefinition}
k_r(r) &=& \sqrt{ g_{rr}(r) \left( - E^2 g^{tt}(r) - \vec{k}^2 g^{\vec{x}\vec{x}}(r) - m^2  \right)} . \, \,
\eea
Therefore, the integration $d \vec{k}$ in 
$\int d\vec{k} k_r(r)$ 
%
can be conducted, and written as  
\begin{eqnarray}
\int d\vec{k} k_r(r) 
&=& \frac{1}{2} \gamma(S^{d-1}) \int d |\vec{k}|^2 \, |\vec{k}|^{d-2}
k_r(r)
\,, \nonumber \\
&=& \frac{c_d}{2} \gamma(S^{d-1})   
\left( \frac{E^2- m^2 f(r)}{f(r)^2} \right)^\frac{d+1}{2} (\rho f(r))^\frac{d}{2}, \nonumber \\
\label{hypergeometric}
\end{eqnarray}
with the factor 
\bea
\label{thedefinitionofcd}
c_d  = \int_0^1 dy y^\frac{d-2}{2} \sqrt{1-y} = \frac{\sqrt{\pi } \Gamma \left(2-\frac{2}{d}\right)}{2 \Gamma
   \left(\frac{7}{2}-\frac{2}{d}\right)} \,, 
\eea
in general dimensions. Here $\gamma(S^{d-1})$ is area of the unit radius
$S^{d-1}$,
$f(r)\equiv -g_{tt} (r)=g^{rr}(r)$ and $\rho \equiv g_{\vec{x} \vec{x} } (r) $. 
The WKB approximation to obtain eq.~(\ref{krdefinition}) is justified if 
\bea
\label{WKBcheck}
\frac{(\sqrt{-g} g^{rr} k_r(r))'}{\sqrt{-g} g^{rr} k^2_r(r)} \ll 1  \,,
\eea
is satisfied. In the near horizon region, which we are most interested in, $g^{\vec{x}\vec{x}}$ 
approach constant, but $g^{tt}$ diverges, therefore we are in the range 
$- E^2 g^{tt}(r) \gg  \vec{k}^2 g^{\vec{x}\vec{x}}(r)$, $- E^2 g^{tt}(r) \gg m^2$, and we can approximate as,  
\bea
\label{approxkrvalue}
k_r(r) \approx \sqrt{- g_{rr}(r) g^{tt}(r)} E = g_{rr}(r) E \,.
\eea
in the gauge $g_{rr}(r) = - g^{tt}(r)$.
Then we see that $\sqrt{-g} g^{rr} k_r(r)$ approach constant value, 
therefore WKB approximation eq.~(\ref{WKBcheck}) \
is satisfied.\footnote{Later we will also consider large momentum region 
$- E^2 g^{tt}(r) \gtrsim - \vec{k}^2 g^{\vec{x}\vec{x}}(r)$, 
there, the WKB approximation is not strictly valid. 
However, we neglect such subtle issues in this paper since we expect that the result will not change much qualitatively.}

So far the argument is for generic dimensions. 
The integration over $\vec{k}$ along (\ref{hypergeometric}) simplifies for the $d=2$ case.  
In that case, it simplifies as 
\bea
\int d\vec{k} k_r(r) &\sim& - \frac{1}{g_{rr}(r) g^{\vec{x}\vec{x}}(r)} 
(k_r(r))^{3/2} 
|^{k_r(r)=0}_{|\vec{k}|=0} \nonumber \\
&=& \frac{1}{g_{rr}(r) g^{\vec{x}\vec{x}}(r)} \left( g_{rr}(r) \left( -E^2 g^{tt}(r) - m^2  \right) \right)^{3/2} \nonumber \\
&\sim& \frac{(- g_{rr}(r) g^{-1}_{tt}(r))^{3/2}}{g_{rr}(r) g^{\vec{x}\vec{x}}(r)} E^3 \nonumber \\
&\sim& (-g_{tt})^{-3/2}  g_{rr}(r)^{1/2} g_{\vec{x}\vec{x}} \, E^3 \,.
\label{2dimresult1}
\eea
where we have neglected the effect of mass term and also irrelevant numerical factors. 
The mass term is suppressed near the horizon compared with $E$ since 
\bea
-E^2 g^{tt}(r) \gg m^2
\label{massrest}
\eea
near the horizon, from which the dominant contributions comes. Given $\int d\vec{k} k_r$, using (\ref{betafreeenergy}), we can obtain the free energy of the 
probe. What 't Hooft pointed out in \cite{'tHooft:1984re} is that this contribution diverges due to the near horizon contribution.

\subsection{Schwarzschild black holes in asymptotic flat space-time in 4 dimension}  

Let us review the 't Hooft's original asymptotic flat Schwarzschild black hole case in $D=4$ dimensions. 
In this subsection, we take the 4d Planck scale set to be unit.  
Note that if we consider the black {\it holes}, then due to the centrifugal force, $\vec{k}$ integration 
is replaced by the $l$ integration and we have additional $r^2$ factor and then, 
using that $g_{rr}(r)$ has single zero at the horizon in Schwarzschild black holes and 
\bea
&& g_{rr}^2 |_{r \approx r_h} \approx (1 - \frac{r_h}{r})^{-2} \,, \quad
 -g_{tt}^{-1}(r) = g_{rr}(r) \,, \quad \CR 
&& g_{ii}|_{r = r_h + h} = r_h^2 + O(h) \,, 
\eea
so eq.~(\ref{2dimresult1}) gives the dominant contributions 
\bea
\int dr dl k_r(r) &\sim& \int dr r^2 g_{rr}(r)^2 E^3  
\sim E^3  ( \frac{L^3}{3}  + \frac{r_h^4}{h} )\,, \quad \quad
\eea
at $L \to \infty$ and $h\to 0$ limit. 
The necessity of factor $r^2$ is clear from the dimensional analysis, $d
\vec{k} = {dl (2 l+1)}/{r^2}$ 
in $D(=d+2)=4$ case.

Therefore, using (\ref{betafreeenergy}) in the black hole case as 't Hooft, after $E$ integration, 
we obtain the free energy per unit volume, 
\bea
F_{\phi}  &\sim& -\frac{1}{h} (\frac{r_h}{\beta})^4 - L^3 \int_m^\infty dE  \frac{ \, (E^2 - m^2)^{3/2}}{e^{\beta E} - 1}  \,.
\eea

This is the free energy of the probe scalar field around the black hole, where it is 
thermal equilibrium with the black hole. This free energy diverges at $h \to 0$. 
However it does not make sense that the probe free energy diverges, and gets bigger than the background 
black brane free energy\footnote{In this case, since it is asymptotic flat space-time, we do not have clear 
holographic interpretation. Later we discuss in more detail about the probe free energy and background 
free energy contribution from gauge/gravity viewpoint in asymptotic AdS.}.  
This forces us to put the cut-off for the minimal values for $h$. 
The black hole entropy is 
\bea
\quad \quad S_{BH} = \frac{r_h^2}{l^2_{p}} 
\, \quad \Rightarrow \quad F_{BH} \sim  \frac{S_{BH}}{\beta} = \frac{r_h}{l^2_{p}} 
\eea
where $l_{p}$ is 4-dimensional Planck scale. The Hawking temperature $T$ is 
\bea
\beta = \frac{1}{T} = {r_h} \,.
\eea
A natural cut-off is given by the background black hole free energy, this gives the cut-off for $h$ as 
\bea
F_{\phi} \lesssim F_{BH} \Rightarrow h \lesssim \frac{l^2_{p}}{r_h}  \sim T\, l_{p}^2 
\eea 
Actually this distance $h$ measured in the coordinate invariant way gives
\bea
\Delta s &=& \int^{r_h + h}_{r_h} dr \sqrt{g_{rr}} \CR 
&\sim& \int^{r_h + h}_{r_h} \sqrt{r_h}  \frac{dr}{\sqrt{r- r_h}} 
= \sqrt{r_h h} = l_{p}
\eea
so it is $O(1)$ in Planck unit. Especially this means that it is {\it independent} on the black hole states (mass, temperature) we consider for the background, and implies rather it is some intrinsic nature of the theory. 

\subsection{Origin of the divergence}
\label{originofdiv}

The essential parts of the 't Hooft's brick wall model is the necessity of the 
near horizon cut-off at $r = r_h + h$, with $h > 0$. 
The origin of the free energy divergence is due to the divergence of the occupation number 
$g(E)$ given in eq.~(\ref{gEdefinition}) and this is associated 
with the infinite volume of the 
deep throat region near the horizon. 
Note that $g(E)$ diverges at finite $E$ means that there are infinitely {\it low energy} spectrum degeneracy. 


As the spatial volume gets bigger in the system, the allowed low energy excitations increase. 
Therefore, given the temperature, we have more entropy. 
This is physically due to the large IR 
regime near the horizon, we can have infinitely 
low energy quantized mode allowed. 
This is exactly happening in the near horizon. 
However, considering the infinite volume factor from 
the radial direction only, we have only a logarithmic
divergence in 
\bea
\int dr k_r \sim \int dr g_{rr} E \sim \frac{E}{T} \log h, \,
\eea
which is not what we have observed.
Actually, in the near horizon region the volume of 
the $\vec{x}$ direction also 
becomes bigger and then
this divergence is enhanced 
after the $\vec{k}$ integration as 
\bea
\int dr d\vec{k} k_r(r) \sim \int dr g^2_{rr} E^3 \sim \frac{E^3}{T^2 h}
\eea 
in the $g_{tt}(r) = g^{rr}(r)$ gauge. 
This is the origin of the $\int dr d \vec{k} k_r \propto 1/h$ divergence.   

In summary, $k_r$ in eq.~(\ref{approxkrvalue}) has $g_{rr}$, but 
the $\vec{k}$ integration, also add another factor $g_{rr}$.  
The $\vec{k}$ integration yields another factor $g_{rr}$ is because of the integration region 
\bea
 0 \, \le\, |\vec{k}|^2 &\le& g_{\vec{x}\vec{x}}|_{r=r_h+h} \left( - E^2 g^{tt}|_{r=r_h+h} - m^2 \right) \CR
&\sim& E^2 g_{\vec{x}\vec{x}} g_{rr}|_{r=r_h+h} 
\sim \frac{T}{h} \,.   
\eea
near the horizon. 
Therefore if we introduce the boundary field theory momentum cut-off $\Lambda$, 
\bea
\label{smallLambdacut-off}
 |\vec{k}|^2 \le \Lambda^2 \ll E^2 g_{\vec{x}\vec{x}} g_{rr}|_{r=r_h+h}  \sim \frac{T}{h} 
\eea
Then, the free energy has divergence only by $F \sim \Lambda^2 \log h$, since 
\bea
 d\vec{k} k_r(r) \sim \int  \Lambda^2 g_{rr} E  
\eea
instead of eq.~(\ref{2dimresult1}), and this gives 
\bea
F \sim  \Lambda^2 T \log h
\eea
using eq.~(\ref{betafreeenergy}).  
It depends on the theory 
whether the theory admits momentum cut-off $\Lambda$ satisfying (\ref{smallLambdacut-off}) or not.

Later we discuss this point for the generic holographic viewpoint. There, since it approaches the 
asymptotic AdS, which is scale invariant theory, it is unnatural to have such a 
momentum cut-off scale $\Lambda$ 
in boundary viewpoint. 

It is pointed out in \cite{Demers:1995dq} that 
a Pauli-Villars regulator induces the similar finite IR cut-off effect.  
In \cite{Demers:1995dq}, a Pauli-Villars regulator field $\phi^{(PV)}$ is introduced, where 
$\phi^{(PV)}$ has opposite sign to the $\phi$ field and has large mass $M$ with $m^2 \ll M^2$. 
Under the same WKB approximation, the regulator field $\phi^{PV}(r) = \phi^{PV}(r)_0 e^{i \int k_r^{(PV)}(r) dr } $ shows the relation which is very similar to 
(\ref{krdefinition}) but $m^2$ is replaced by $M^2$ as 
\bea
\label{kMdefinition}
k^{(PV)}_r(r) &=& \sqrt{ g_{rr}(r) \left( - E^2 g^{tt}(r) - \vec{k}^2 g^{\vec{x}\vec{x}}(r) - M^2  \right)} . \quad \, \,
\eea
This $k^{(PV)}$ approaches $k_r$ for $\phi$ given by (\ref{krdefinition}) at 
\bea
 - E^2 g^{tt}(r) \gg M^2  \,, \quad   \vec{k}^2 g^{\vec{x}\vec{x}}(r) \gg M^2 \,.
\eea
These are equivalent to 
\bea
r-r_h \ll \frac{T}{M^2} \,,  \quad \vec{k}^2   \gg M^2 \,.
\eea
However as we have seen in this subsection, 
the divergence of the free energy is exactly originated by the many modes in 
these parameter region. Therefore a Pauli-Villars regulator $\phi^{(PV)}$ with mass $M$ 
plays the same role as introducing the cut-off $h$ with 
\bea
\label{hTMsqrelation}
h \sim \frac{T}{M^2} \,.
\eea
Since the local temperature at $r = r_h + h$ is
\bea
T_{local} = \frac{T}{\sqrt{g_{tt}|_{r=r_h +h}}} \sim M \,.
\eea 
this regulator $\phi^{(PV)}$ removes the degrees of freedom 
\bea
\label{divergingmodes}
E \gtrsim M \,, \quad \vec{k}_{local}^2 \gtrsim M^2 \,, 
\eea
where we define $\vec{k}_{local}^2 =g^{\vec{x} \vec{x}}|_{r=r_h +h} \, \vec{k}^2$,
so that the divergent contribution is cut-off. 
In terms of $h$, this is 
\bea
\label{divergingmodes2}
E \gtrsim \sqrt{\frac{T}{h}} \,, \quad \vec{k}_{local}^2 \gtrsim \frac{T}{h} \,.
\eea

It is clear that any regulator, other than above Pauli-Villars which removes the 
local modes (\ref{divergingmodes}), plays the same role as introducing nonzero cut-off $h$ to remove 
the modes (\ref{divergingmodes2})\footnote{%
Such an removal of UV degrees of freedom should be considered as a result
of the strong graviatational interactions near the horizon. 
Note that this is a UV cut-off in the bulk gravitational theory,  which corresponds to  
the IR cut-off in the holographic dual boundary theory through the UV/IR relation \cite{Peet:1998wn}.
}. 
And if $M$ is Planck scale, the regulated free energy $F_{\phi}$ becomes the same order to the 
Gibbons-Hawking black hole free energy. Regarding the black hole entropy as this $F_{\phi}$, 
and replacing the black hole by the brick wall is the original idea of 't Hooft's brick wall model 
for the black holes \cite{'tHooft:1984re}. However in this paper, 
we interpret this divergence rather as an evidence of the composite nature of the any 
probe fields 
in holographic dual field theory. 
Since in dual field theory, we can have clear interpretation of the probe scalar fields as 
new degrees of freedom, it is more natural not to regard $F_{\phi}$ as the background 
black hole entropy $F_{BB}$. But for a moment, we keep this issues aside and continue the 
universal nature of the model. Then later in \S IV, we discuss dual interpretations.

\section{Universality of brick wall models}
In previous section, 
we have reviewed the original 't Hooft's brick wall model. 
In this section, 
we first consider the brick wall model for the 4-dimensional 
Schwarzschild black brane in asymptotic anti-de Sitter and see 
the same results hold. In some sense, this is expected since the brick wall 
model is determined by the near horizon nature and 
both Schwarzschild black brane in asymptotic flat space-time and in asymptotic 
AdS has the same near horizon. 

Then, we will study more generic setting of the brick wall properties. 
We will now generalize the previously studied nature of the universality of the brick 
wall in more generic black brane and see how universal it is.

In order to check this universality, we consider more generic situation 
where 
the background black brane shows various exotic nature, some of which  
includes the recently studied 
Lifshitz black branes \cite{Taylor:2008tg, Goldstein:2009cv, Goldstein:2010aw}, 
hyperscaling violating black branes \cite{Charmousis:2010zz, Iizuka:2011hg}.  
There, the background scalar/gauge fields do not take a constant VEV but rather takes 
non-trivial dynamics. 
We will conduct the analysis for both neutral scalar and 
also the charged scalar on non-trivial background, with and without minimal coupling assumption.  
We will see its universality under the mild conditions, probe field free energy always diverge and 
its equality with the background exotic black brane yield the Planck scale. 
We will also check for the model on the probe brane degrees of freedom. 
Then we check similar things this for fermions.

\subsection{Schwarzschild black brane in AdS$_4$} 
Let's first consider the Schwarzschild black brane in AdS$_4$. 
In this case, the metric is 
\bea
ds^2 &=& - f(r) dt^2 + \frac{1}{f(r)} dr^2 + r^2 d \vec{x}^2 \,, \CR 
f(r) &=& r^2 - r_h^2 \,,
\eea
Here we have set the AdS curvature scale to be unit. 

Since $g_{rr}(r)$ has single zero at the horizon  
and we have 
\bea
&& g_{rr}^2 |_{r \approx r_h} \approx r_h^{-2} (r - {r_h})^{-2} \,,\\
&& -g_{tt}^{-1}(r) = g_{rr}(r) \,,\quad  g_{\vec{x} \vec{x}}|_{r = r_h + h} = r_h^2 + O(h)  \,.\quad
\eea

Therefore after $r$ integration using eq.~(\ref{betafreeenergy}), we have free energy divergence as $1/h$,
\bea
\int dr d\vec{k} k_r(r) &\sim& \int dr g_{rr}(r)^2 g_{\vec{x} \vec{x}}(r_h) E^3 \nonumber \\
&\sim & E^3 (\frac{1}{L} + \frac{1}{h})
\eea
where we have used $g_{rr}(r)^2 \to 1/r^4$, $g_{\vec{x} \vec{x}}(r) \to r^2$ at $r \to L \sim \infty$.  
At the $h \to 0$ limit, this free energy for the probe scalar field in black brane background diverges. 
And we obtain the free energy per unit volume
\bea
F_{\phi}  &=& -\frac{1}{h} \frac{1}{\beta^4} -  \frac{1}{L} \int_m^\infty dE  \frac{ \, E^3}{e^{\beta E} - 1}  \,.
\eea

In the black brane case, the background black brane entropy is proportional to the spatial dimension 
volume, so the free energy per unit volume is 
\bea
S_{BB} = \frac{1}{l^2_{p}} \, g_{\vec{x} \vec{x}}|_{r = r_h}  \, \quad \Rightarrow \quad F_{BB} \sim  \frac{S_{BB}}{\beta} = \frac{1}{l^2_{p}} \frac{r_h^2}{\beta} 
\eea
Therefore, we have cut-off 
\bea
h \sim \frac{l^2_{p}}{r_h^2\beta^3} \,, 
\eea
Since we have temperature in terms of $r_h$ as 
\bea
\beta = \frac{1}{r_h} \,.
\eea
\footnote{To see this, note that our metric has in Euclidean signature, 
$ds^2 = r_h(r-r_h)d\tau^2 + \frac{dr^2}{ r_h(r-r)}$. By coordinate changes $\tilde{r} = \int  dr/\sqrt{r_h(r - r_h)} \sim \sqrt{(r - r_h)/r_h}$, we have $ds^2 =  \tilde{r}^2 r_h^2 d\tau^2 + {d\tilde{r}^2}$. This gives 
$\beta \sim 1/r_h$, or $T \sim r_h$.}
Therefore the cut-off $h$ is 
\bea
h \sim {r_h}{l^2_{p}} =T \, {l^2_{p}}  \,.
\eea
The distance $h$ measured in the coordinate invariant way (the invariant distance)
\bea
ds &=& \int^{r_h + h}_{r_h} dr \sqrt{g_{rr}} \CR 
&\sim& \int^{r_h + h}_{r_h} \frac{1}{\sqrt{r_h}} \frac{dr}{\sqrt{r- r_h}} 
= \sqrt{\frac{h}{r_h}} \sim l_{p}.
\eea

\subsection{Lifshitz and Hyperscaling violating black brane in 4 dimensions}

\subsubsection{Background near horizon solution}
To study more generic setting, let's consider the following dilaton
Maxwell and gravity model \cite{Charmousis:2010zz, Iizuka:2011hg}. 
\bea
\label{action}
S = \int d^4\!x \sqrt{-g} \left\lbrace R - 2 (\nabla \phi)^2 -e^{2 \alpha \phi} F_{\mu \nu} F^{\mu \nu} - V_0 e^{2 \delta \phi}\right\rbrace . \quad
\eea
This theory allows more exotic black brane solutions. 
We consider the case $V_0<0$. We are assuming that this is the effective theory valid 
near the horizon, which smoothly 
connect to the UV region where we have constant dilaton and the metric approaches AdS \footnote{This can be achieved by setting the potential as $V(\phi) = 2 V_0 \cosh 2 \delta \phi$, for example, and $\phi \to 0$ in UV.}.
Taking the metric ansatz, 
\bea
\label{ansatz1}
 ds^2 = - a(r)^2 dt^2 + \frac{ dr^2}{ a(r)^2 } + b(r)^2 (dx^2 + dy^2) \,, 
 \eea
The Maxwell equations are satisfied by 
\bea
\label{gf}
 F = {Q_e \over f(\phi) b^2 }  dt \wedge dr \,,
\eea
and the remaining equations of motion can be conveniently expressed in terms of an effective potential \cite{Goldstein:2005hq}
\begin{equation}\label{Veff}
V_{{eff}} = {1 \over b^2} \left( e^{-2 \alpha \phi} Q_e^2  \right) + {b^2 V_0  \over 2} e^{2 \delta \phi},
\end{equation}
as
\bea 
&& (a^2 b^2  \phi')'  =   {1 \over 2} \partial_\phi V_{{eff}} \,, \quad
(a^2 b^2)'' =  -2 V_0 e^{2 \delta \phi} b^2 \,, \CR 
&& {b'' \over b} =  -  \phi'^2  \,, \quad  
 a^2 b'^2 + {1 \over 2} {a^{2}}' {b^{2}}' =   a^2 b^2 \phi'^2 - V_{{eff}}. \quad
\eea
Solving equations of motions with the ansatz,  
\bea
a = C_a r^{\gamma} \,, \hspace{10mm} b= r^\beta \,, \hspace{10mm} \phi = k \log{r} \,, 
\eea
we obtain 
\bea\label{case1}
&& \beta = { (\alpha+\delta)^2 \over 4 + (\alpha+\delta)^2}   \,, \,\,  
\gamma = 1 -{ 2 \delta (\alpha+\delta) \over 4 + (\alpha+\delta)^2}  \,, \CR 
&& k = - { 2 (\alpha+\delta) \over 4 + (\alpha+\delta)^2}   \,, \,\,  
Q_e^2 = -V_0{ 2 - \delta (\alpha+\delta) \over 2 \left( 2 + \alpha (\alpha+\delta) \right)}  \,,\CR
\label{case11} 
&& C_a^2  = -V_0 {\left( 4 + (\alpha+\delta)^2 \right)^2 \over 2 \left(2 + \alpha (\alpha+\delta) \right) \left( 4 + (3 \alpha-\delta) (\alpha+\delta) \right)} \,,   \quad \quad 
\eea

Finite temperature solution is obtained from the fact that 
the equations of motion continue to hold under the following shift 
\bea
a^2 b^2 \to a^2 b^2 + C r \,,\quad  b^2 = \mbox{fixed} \,, \quad \phi = \mbox{fixed} 
\eea
where $C$ is any constant, which we will set $C=C_a^2 r_h^{2 \beta + 2 \gamma -1}$, for convenience. 
This allows us to obtain the finite temperature solution 
\bea
\label{onepa}
a^2=C_a^2r^{2\gamma}\left(1-\left({r_h\over r}\right)^{2\beta+2\gamma-1}\right)
\eea
In the near horizon, this becomes 
\bea
\label{nha}
a^2(r) \simeq (2\beta+2\gamma-1) \left( {r-r_h}  \right) r_h^{2\gamma-1} C_a^2 \,.
\eea
The temperature of the black brane solution is given by 
\bea
\label{exoticTrhrelation}
T \sim C_a^2 r_h^{2 \gamma -1} \,.
\eea

\subsubsection{Scalar fields on these background}

Let's consider the minimally coupled scalar field in this set-up. 
Quite analogous way to the \S II, we obtain the wave equations using the WKB approximation. 
The argument of \S II.A, II.B till equation (\ref{2dimresult1}) still holds as far as $d=2$.  
Since the metric is different we have,
\bea
g_{rr} = \frac{1}{a(r)^2}
\eea
and, 
\bea
\int dr d\vec{k} k_r(r) &\sim& \int dr \frac{b(r)^2}{a(r)^4} E^3 \CR
&\sim&  \frac{r_h^{2\beta}E^3}{C_a^4 r_h^{4 \gamma -2 }h} + (L\, \mbox{dependent part})  \quad
\eea
where for the first relation, we used (\ref{2dimresult1}).
So the free energy of the scalar $F_{\phi}$ per unit volume is 
\bea
F_\phi &\sim& \int \frac{dE}{1 - e^{-\beta E}} dr d\vec{k} k_r(r)  \CR 
&\sim& \frac{r_h^{2\beta}}{C_a^4 r_h^{4 \gamma -2 } h \beta^4} + (L\, \mbox{dependent part})  \,.
\eea

On the other hand, the generalized Lifshitz black brane temperature is $T \sim C_a^2 r_h^{2 \gamma -1}$, 
and the entropy of the black brane is 
\bea
S_{BB} \sim \frac{1}{l^2_{p}}\, r_h^{2\beta}
\eea
so the free energy is 
\bea
F_{BB} \sim T S_{BB} \sim\frac{1}{l^2_{p}}  C_a^2 r_h^{2 \gamma + 2 \beta -1} \,.
\eea
Comparing this with $F_\phi$, we have cut-off of $h$ as 
\bea
 F_{BB} &\lesssim& F_{\phi} \,,\\ 
\Rightarrow 
h  &\lesssim& \frac{l^2_{p} }{\beta^4 C_a^6 r_h^{6 \gamma - 3}}  \sim 
{C_a^2 r_h^{2 \gamma -1}}  l^2_{p}  \sim T l^2_{p}  \,,
\eea 
Then the invariant distance is  
\bea
\Delta s &=& \int_{r_h}^{r_h + h} dr \sqrt{g_{rr}(r)}  
\sim
\frac{\sqrt{h}}{  
C_a r_h^{(2 \gamma -1)/2}}  \sim l_{p} \,.
\eea
Again, it is Planck length in terms of invariant distance, and independent on $\beta$ and $\gamma$ which characterize IR.

\subsection{Generic dimension argument}

In this subsection, 
we will show that the universality we have seen is from
the Rindler like geometry.
Especially, we will extend the previous results to
the Schwarzschild black brane in AdS$_D$ space-time. 
In this subsection, we keep the all length scale explicit. However we neglect irrelevant 
numerical coefficients. 

The geometry we consider here is the Rindler like geometry
\begin{eqnarray}
 ds^2&=& -f (r) dt^2 +\frac{1}{f(r)} dr^2 +\rho d\vec{x}^2, \, \CR 
 f(r) &=&T (r-r_h),
\end{eqnarray}
for the near horizon, 
where $t_H, r_h, \rho = \rho(r_h)$ is constant and $\vec{x}$ represent the $d=D-2$
dimensional flat space with compactification such that
$\int d \vec{x}=V$. $T$ is the Hawking temperature. 
This geometry is the near horizon geometry of, 
generic black branes\footnote{For example, for 
the Schwarzschild black brane  in $AdS_D$ space-time, we have 
$T=r_h/(R_{AdS})^2$ as the Hawking temperature
and $\rho=(r_h/R_{AdS})^2$. In more generic Lifshitz or Hyperscaling violating geometry, 
this power is more exotic as is given in eq.~(\ref{exoticTrhrelation}).}.
Then again, using the WKB approximation, we have 
\begin{eqnarray}
\int d\vec{k} k_r(r) 
&=& \frac{1}{2} \gamma(S^{d-1}) \int d |\vec{k}|^2 \, |\vec{k}|^{d-2}
k_r(r)
\,, \nonumber \\
&=& \frac{c_d}{2} \gamma(S^{d-1})   
\left( \frac{E^2+ m^2 f(r)}{f(r)^2} \right)^\frac{d+1}{2} (\rho f(r))^\frac{d}{2}, \nonumber \\
\end{eqnarray}

The factor $c_d$ is given in eq.~(\ref{thedefinitionofcd}), and it is order one irrelevant numerical factor, 
so we have 
\begin{eqnarray}
F_\phi &\sim& V \int dr \int \frac{dE}{1-e^{-E/T}} 
\left( \frac{E^2}{f(r)^2} \right)^\frac{d+1}{2} (\rho f(r))^\frac{d}{2} 
\nonumber \\
& \sim & V \rho^\frac{d}{2} (T)^{d+2} \int_{r_h + h} dr (f(r))^{-\frac{d+2}{2}} \nonumber \\
&\sim&  V \rho^\frac{d}{2} (T)^\frac{d+2}{2} h^{-\frac{d}{2}},
\label{Fphishdependencebydim}
\end{eqnarray}
again we have neglected the numerical factor and $m^2$ term which 
is not the leading contribution near the horizon since $f(r) \to 0$ at the horizon. 

Note that since the mass term is in the end negligible due to the warped factor implies that 
the argument above is not unchanged for non-minimally coupled scalar fields. This is because 
the effects of non-minimal coupling, like curvature couplings, are included to ``renormalize'' the mass term. 
However in the end the mass term is not dominant compared with the other terms, so 
above argument hold more generic situations.

On the other hand, the BH entropy of this geometry is
$S_{BB} \sim V \rho^\frac{d}{2}/ (l_p)^d$. From this, we have
\begin{eqnarray}
 F_{BB} \sim T V \rho^\frac{d}{2} \frac{1}{(l_p)^d}.
\end{eqnarray}
Assuming $F_\phi \sim F_{BB}$,
we have
\begin{eqnarray}
 h \sim T l_p^2 ,
\end{eqnarray}
and the distance $h$ measured in the coordinate invariant way 
\bea
ds &=& \int^{r_h + h}_{r_h} dr \sqrt{g_{rr}} \CR 
&\sim& \int^{r_h + h}_{r_h}  \frac{dr}{\sqrt{(r -r_h ) \,T }} 
\sim \sqrt{\frac{h}{T}} \sim l_p,
\eea
which is universal as before. 

We find the local temperature at $r=r_h+h$ is
\begin{eqnarray}
 T_{local}  = \frac{T}{\sqrt{f|_{r=r_h +h}}} \sim \frac{1}{l_p},
\label{localTT}
\end{eqnarray}
which is also universal.

Actually, the local temperature can not be much larger than the Planck
scale in our analysis, otherwise the quantum correction will be needed.

The region of the momentum $\vec{k}$ which give the 
dominant contributions for this $h$ is 
\bea
\label{localkk}
|\vec{k}| \sim \frac{\sqrt{\rho}}{\sqrt{T h}} \, E \sim \frac{\sqrt{\rho}}{l_p}  \sim \left( {\frac{S_{BB}}{V}} \right)^{1/d} \,.
\eea
which is the order of the entropy density\footnote{Eq.~(\ref{localTT}) and (\ref{localkk}) imply that
the quantum gravity effects can not be neglected near $r=r_h+h$. 
The role of brick wall is to remove the degrees of freedom for 
the quantum gravity for local temperature/momentum above a cut off scale $\sim 1/l_p$. 
This can be done also by introducing the regulator which effectively cut off 
the geometry for $r \lesssim r_h+h$ 
without the explicit brick wall as \cite{Demers:1995dq}.}.

Finally, let us consider what happen if we assume the 
free energy of the background black brane and the one for the probe has 
different $N$ scaling, as 
\bea
N^\delta F_\phi \sim F_{BB}  \sim N^\alpha \,.
\eea
For example, usual $N_c = N$ D3-branes with $N_f$ flavor branes system 
gives 
\bea
F_{BB} &=& N^2 T^{d+1} f_{BB}(g_{YM}, T) \,, \quad \CR  
F_\phi &=& N_f N T^{d+1} f_{\phi}(g_{YM}, T) \,,
\eea
in our limit, where $g_{YM}$ and T are fixed and $N$ is very 
large and $f_{BB}$ and $f_\phi$ are some functions. 
In this case, $\alpha = 2$, $\delta = 1$\footnote{$N$ D3-branes have $N^2$ degrees of freedom due to adjoint matrices, and $N_f$ D7-branes 
have $N$ degrees of freedom due to fundamental quarks. This is consistent with the 
famous 3/4 factor difference between $\lambda \to \infty$ limit and $\lambda \to 0$ limit \cite{Gubser:1996de,Gubser:1998nz}.}. However 
in more generic D$p$-branes at non 't Hooft limit and also branes in M-theory, $\alpha$ 
and $\delta$ varies, generically. 
We require that the free energy of the scalar field to be probe field yields, 
\bea
0 \le \delta \le \alpha \,.  
\eea
Below we will keep only $l_p$ as a dimension-full parameter.
Since 
\bea
\label{fphi}
F_{BB} \propto G_N^{-1} \propto l_{p}^{-d} \,, \quad F_{\phi} \propto h^{-\frac{d}{2}} \,,
\eea
from eq.~(\ref{Fphishdependencebydim}), 
this means 
\bea
\label{lphNrelation}
\l_p \sim N^{-\frac{\alpha}{d}} \,, \quad \mbox{and} \quad h \sim N^{\frac{2(\delta-\alpha)}{d}} 
\sim (l_p)^{\frac{2(\alpha-\delta)}{\alpha}} \,.
\eea

Note that this $l_p$ is defined from the relation $ G_N \propto l_{p}^{d}$, so $l_p$ is Planck length 
in the bulk AdS$_{d+2}$. 
Note also that the IR cut-off $h$ in eq.~(\ref{lphNrelation}) is written {\it not} in terms of invariant distance 
but rather in terms of $r$. 
$h \sim l_p$ if $F_{BB} \sim (F_\phi)^2$. 

Indeed, 
for the large black hole in the AdS coming from 
typical D$3$-D$7$ system in holographic QCD setting 
in the high temperature phase,
the $F_{BB}$ is expected to be scale like $N^2$
and the $F_\phi$ for the probe brane theory scale like $N$, 
%
since we have $N$ D$p$-branes for $O(N^2)$ d.o.f. from adjoint representation and $O(1)$ probe D$p'$-branes for $O(N)$ d.o.f. from fundamental representation of $SU(N)$. 

In the string theory context, usually 
there are Kaluza-Klein (KK) modes for the probe field.
If the mass of a KK mode satisfies 
\begin{eqnarray}
 m^2 \ll \frac{T}{h} \,,
\end{eqnarray}
then the KK modes give the same contribution to the free energy 
as the massless modes. 
The number of such modes will be $ \left({ \frac{T}{h}} \right)^{\frac{\gamma}{2}}$,
where $\gamma$ is, typically, a dimension of the compactified manifold. 
Here the scale of the manifold was assumed to be $O(1)$. 
Thus including the KK modes, we have 
\bea
F_{\phi} \propto h^{-\frac{d}{2}} \left( \frac{T}{h} \right)^{\frac{\gamma}{2}}
\,,
\eea
instead of (\ref{fphi}) and 
\bea
h \sim (l_p)^{\frac{2(\alpha-\delta)}{\alpha} \frac{d}{d+ \gamma}} \,.
\label{hkk}
\eea
Below we will asumme that there are no KK-modes
for notational simplicity
because the essential discussions are not altered by replacing 
(\ref{lphNrelation}) to 
(\ref{hkk}).

\subsection{Charged scalar hair model}

The effects of background charges for the black branes, which are dual to the chemical potential, 
are modification of the bulk momentum into covariant momentum. Therefore only modification due to the 
background flux are WKB approximated equation (\ref{krdefinition}) is replaced by  
\bea
\label{krmodifiedduetoflux}
k_r(r) = \sqrt{g_{rr}(r) \left( - g^{tt}(r) ({\cal{E}})^2  -  g^{ii}(r) (\tilde{k}_i)^2 -  m^2 \right)} 
\eea
where
\bea
{\cal{E}} \equiv E + q A_0 \,, \quad {\tilde{k}}_i \equiv k_i - q A_i
\eea
The $\vec{k}$ integration in $d \vec{k} k_r(r) $ is not modified since the integration range is still 
the range where the inside of the square-root in (\ref{krmodifiedduetoflux}) is positive, and 
$A_i$ gives just the integration parameter shift. 
$A_0$ approaches zero at the horizon\footnote{If $A_0$ does not vanish at the horizon, 
by analytically continuing it to the Euclid signature, it yields non-trivial Wilson loop $W = e^{i \int A_0 d\tau} = 
e^{i 2 \pi A_0}$ 
on the tip of the $(\tau, r)$ cigar geometry. But since this is contractable circles, it gives contradiction.}. 
Therefore the effects of the flux do not modify essentially our 
the brick wall free energy divergence properties.  

We have been considering minimal coupling. However as we have seen, the effect of the 
mass term is negligible for the brick wall argument, since $-E^2 g^{tt} \gg m^2$. 
Non-minimal couplings make the mass term position dependent, 
so this is effective mass changes. However as far as its value 
near the horizon does not diverge, still the mass term is negligible due to $- g^{tt} \to \infty$ near the horizon 
and our brick wall argument is not modified.

\subsection{Probe brane model}
\label{probebranemodelsubsection}

We have seen the universal nature of the probe scalar fields. 
Note that we have never identified the bulk microscopic (or string theory) origin of the scalar fields. 
The scalar fields could be simple Kaluza-Klein mode from higher dimension compactified to 
lower dimensional AdS, or it could comes from some moduli fields in Calabi-Yau compactifications. 
It could be either closed string mode or open string mode. For example, in the D3-D7 system,  
open strings on the probe D7-brane, which are scalar ``meson'' fields, obey the similar wave equation 
as (\ref{wavewaveeq}), from the probe DBI action in the $l_s \to 0$ limit. 

Generically, 
if we consider introducing probe branes in the background, 
we will have some probe fields in the background at least in low energy limit. 
The contents of fields and the interactions are fixed by the type of probe branes and 
how they embed. 
However, on the black hole background geometry which has 
an infinite throat, the nature that low energy continuous spectrum 
appear for any fields on the probe branes are universal as far as 
probe brane are touching the black hole horizon. 
Thus, the above argument can be essentially applied for any probe fields. 

Even for probe branes which does not cover the whole AdS space time,
for example D8-branes in D4-D8, we expect that
the branes will intersect the horizon of the black hole generically. 
In fact, for D8-brane in D4-D8 system see \cite{Aharony:2006da} for high finite temperature solutions where 
probe brane intersect with horizon.

Actually, if we consider larger and larger black holes in AdS,
the branes eventually intersect the horizon and once horizon and probe 
brane intersect, essentially the same argument shows the probe meson 
free energy diverges.

\subsection{Argument for fermions}

We shall explicitly calculate minimally coupled Dirac fermion with the background gauge field.
We work in the Lorentzian  signature, while the Euclidean signature (which we employed in
the derivation) can be easily obtained by an analytic continuation.
First, we derive the Dirac equation. The fermion action in the bulk is
\begin{eqnarray}
S_{\rm fermion} = \int \! d^{3+1}x \; \sqrt{-g} \, i \,
\left[\, \bar{\psi} \Gamma^M D_M \psi - m\bar{\psi}\psi \, \right].
\end{eqnarray}
Here the Dirac operator is $D_M = \partial_M + \frac{1}{4} w_{abM}\Gamma^{ab} - i q A_M$. 
The definition of the
Gamma matrices in the local Lorentz frame are
\begin{eqnarray}
\Gamma^{\underline{z}} \equiv \left(
\begin{array}{cc}
{\bf 1}_2 & {\bf 0}_2 \\
{\bf 0}_2 & {\bf -1}_2
\end{array}
\right),
\quad
\Gamma^{\underline{\mu}} \equiv \left(
\begin{array}{cc}
{\bf 0}_2 & \gamma^\mu \\
\gamma^\mu & {\bf 0}_2
\end{array}
\right)
\end{eqnarray}
with $\gamma^0 \equiv i\sigma_3$, $\gamma^1 \equiv \sigma_1$, and $\gamma^2\equiv -\sigma_2$, 
where $\sigma_1, \sigma_2, \sigma_3$
are the Pauli matrices. We follow the notation of 
\cite{Liu:2009dm} and \cite{Iizuka:2011hg} except for the assignment of $\gamma^\mu$
(this difference is necessary to see the diagonalization as for fermion components, see below).
The notation for the indices are: $M=0,1,2,z$, and $\mu=0,1,2$.

Writing the 4-component fermion as 
\begin{eqnarray}
\psi \equiv \left(
\begin{array}{c}
\psi_+
\\
\psi_-
\end{array}
\right), \quad 
\psi_\pm \equiv (-g g^{rr})^{-1/4} \phi_{\pm},
\end{eqnarray}
where $\phi_\pm$ is a two-spinor,
the Dirac equation is 
\begin{eqnarray}
\sqrt{\frac{g_{ii}}{g_{rr}}} (\partial_r \mp m \sqrt{g_{rr}})\phi_\pm
= \mp i K_\mu \gamma^\mu \phi_{\mp},
\label{dirack}
\end{eqnarray}
with $K_0 \equiv -i \sqrt{\frac{g_{ii}}{-g_{tt}}} (\partial_0 - i q A_0)$ and $K_i \equiv -i (\partial_i - i q A_i)$ 
with $i=1,2$.

The Dirac
equation (\ref{dirack}) is a coupled equation of $\phi_+$ and $\phi_-$, but one can eliminate
one of them. Bringing (\ref{dirack}) into the following form formally,
\begin{eqnarray}
( - i K_\mu \gamma^\mu)^{-1}\sqrt{\frac{g_{ii}}{g_{rr}}} (\partial_r - m \sqrt{g_{rr}})\phi_+
= \phi_-,
\label{phi-so}
\\
( i K_\mu \gamma^\mu)^{-1}
\sqrt{\frac{g_{ii}}{g_{rr}}} (\partial_r + m \sqrt{g_{rr}})\phi_-
=  \phi_+,
\end{eqnarray}
we can combine these to eliminate $\phi_-$, to have
\begin{eqnarray}
\left[
( i K_\mu \gamma^\mu)^{-1}
\sqrt{\frac{g_{ii}}{g_{rr}}} (\partial_r + m \sqrt{g_{rr}}) ( - i K_\mu \gamma^\mu)^{-1} 
\right. \CR
\left.
\sqrt{\frac{g_{ii}}{g_{rr}}} (\partial_r - m \sqrt{g_{rr}})
\right]\phi_+
= \phi_+.
\end{eqnarray}
This is a second order differential equation for a two-spinor $\phi_+$, 
so we generically have four independent solutions. Without magnetic field, 
given $E$, four states are degenerate and they correspond to 
spin \{up and down\}, and \{normalizable and non-normalizable\} modes.

Under the WKB approximation, this equation becomes 
\begin{eqnarray}
\left[ g_{ii}
\left( g^{rr} \partial^2_r - m^2 \right)
\right]\phi_+
= ( K_\mu \gamma^\mu)^{2}  \phi_+ \,,
\end{eqnarray}
and with the ansatz $\phi_+ = e^{- i E t + i \int dr k_r(r) + i \vec{k} \vec{x}}$, we obtain 
\bea
&& \left[
g_{ii}
\left( g^{rr} k^2_r + m^2 \right)
\right] 
= -( K_\mu \gamma^\mu)^{2} \nonumber \\
&=& - \left(- \sqrt{\frac{g_{ii}}{-g_{tt}}} (E + q A_0) \gamma^0 + (k_i - q A_i) \gamma^i  \right)^2 \nonumber \\
&\sim& \frac{g_{ii}}{-g_{tt}}(E + q A_0)^2 - (k_i - q A_i)^2 \,,
\eea
so we obtain for $k_r$ as 
\bea
k_r(r) = \sqrt{g_{rr}(r) \left( - g^{tt}(r) ({\cal{E}})^2  -  g^{ii}(r) (\tilde{k}_i)^2 -  m^2 \right)} 
\eea
where
\bea
{\cal{E}} \equiv E + q A_0 \,, \quad {\tilde{k}}_i \equiv k_i - q A_i.
\eea
Since fermions are self-conjugate, taking Dirichlet boundary condition for Dirac fermion 
at $r = r_h + h$ is too restrictive. 
Instead, we have to take an appropriate boundary condition such that 
only half of the degrees of freedom are fixed and also that 
energy flow at that boundary vanishes. 
By taking any such appropriate boundary conditions
for the fermions, the same argument holds for fermions. 
Again, the difference from the background flux $A_0$, $A_i$ does not change the result.   
$k_i$ integration is the same, $A_i$ gives just a shift of the integration. 
Near the horizon, $A_0$ vanishes, therefore for the brick wall argument, 
this shows that essentially the same nature holds in fermion case with charged background.

\section{Holographic interpretation of the brick wall}

In the brick wall model, 
the Dirichlet boundary condition is imposed 
on the probe scalar field: $\phi(r=r_h+h)=C_{r_h+h}$. 
The scalar field is assumed to be free.
The boundary is on $r=r_h+h$ with $h>0$,
i.e. outside the horizon.
We first discuss the various IR boundary condition. 

\subsection{Various IR boundary conditions}

In the Lorentzian AdS/CFT context,
we usually take the ingoing boundary condition at 
the horizon since classically horizon keep absorbing objects. 
In fact, this choice of boundary condition allows us to evaluate the retarded Green function.
However, it is also possible to take the Dirichlet boundary 
condition at least for probe fields once we go slightly away from the
horizon\footnote{We should stress that we do not think these boundary
conditions are really realized near the horizon. 
Instead, they are effective descriptions 
for quantum gravity,  
which represent where the semi-classical picture 
breaks down.
As we will see later, this breakdown is related to the finite $N$
effects.
Thus, the usual computation without the brickwall (using the in-falling 
boundary condition) with the $1/N$ expansion 
may be consistent with the brickwall model.}. 
One can say that our boundary condition is on the ``stretched horizon'' 
\cite{Susskind:1993if}. Since it is a time-like 
surface, it is natural to consider either Dirichlet or Neumann boundary condition.
Here we summarize some obvious properties of the Dirichlet 
(or Neumann) and ingoing boundary conditions.

Near the boundary at $r \approx r_h+h$, with $h >0$, the equations of motion allow 
 two independent approximated solutions with fixed $E$ are
$\phi \sim e^{i (E t+k_r(r_h) r)} $ and $\phi \sim e^{i (E t-k_r(r_h) r)} $, 
where $k_r(r_h)$ is the value of $k_r$ in (\ref{krdefinition}) 
evaluated at $r = r_h$, with 
$\mbox{Re}(E/k_r(r_h)) \geq 0$. 
From the action principle, we know not only the equation of motion but
 also what is a consistent boundary condition.  
If we take the ingoing boundary condition
which means $\phi \sim e^{i (E t+k_r x)} $,
it does not make the boundary term, $\phi \, \partial_r \phi|$, vanish. 
Thus generically in such case the energy is not conserved through the 
boundary term and the amplitudes of scalar fields decay/amplify as time evolution. 
This is reflected to 
$\mbox{Im}(E) \neq 0$. 
Instead, if we take the Dirichlet boundary condition $\phi = C_{r_h + h}$ at $r = r_h + h$ with 
constant $C_{r_h + h}$, 
it makes the boundary condition satisfied only with $C_{r_h + h}=0$. 
In this case, the eigenmodes proportional to $e^{i E t} \sin (k_r(r_h) (r - r_h))$, where
$t$ is the static time outside the horizon, 
and we have real $E$, $\mbox{Im}(E) = 0$. 

These can be seen explicitly 
from the bulk equation of motion as follows 
\begin{eqnarray}
0&=& \int_{r_h+h}^\infty \sqrt{-g} 
 \bar{\phi} \left(
\frac{1}{\sqrt{-g}} \partial_r (\sqrt{-g} g^{rr} \partial_r \phi ) \right. \CR 
&& \quad \quad \quad  \quad \quad \quad \left. -E^2 g^{tt} \phi - k^2 g^{\vec{x}\vec{x}} \phi - m^2 \phi
\right) \nonumber \\ 
&=& \int_{r_h+h}^\infty 
\left(
-(\partial_r \bar{\phi}) (\sqrt{-g} g^{rr} \partial_r \phi ) 
+\sqrt{-g} 
\left(-E^2 g^{tt}  \right. \right. \CR 
&& \left. \left. - k^2 g^{\vec{x}\vec{x}}  - m^2  \right) \bar{\phi}\phi \right) +(\sqrt{-g} g^{rr} \bar{\phi} \partial_r \phi)_{r=r_h+h} \,. \quad \quad 
\end{eqnarray} 
Here we consider the normalizable modes, 
i.e. $\phi \rightarrow 0$ as $r \rightarrow \infty$ so that the UV AdS boundary contribution vanishes, {\it i.e.,} 
net energy flow at infinity.  
\footnote{We restrict the mass of the scalar field such that it has a normalizable
and a non-normalizable modes for simplicity.}
For $C_{r_h + h}=0$, taking the imaginary part of this,
we find $\mbox{Im}(E^2)=0$ and $E^2 \geq 0$.\footnote{For $C_{r_h + h} \neq 0$, 
the charge density is not conserved at the boundary
in general.}  
This means that there is no tachyonic modes,
even though there can be modes with $E^2 \ll \vec{k}^2$.
For a general boundary condition,
the imaginary part of this equality relation gives 
\bea
\mbox{Im}(E^2)  \int_{r_h+h}^\infty 
\sqrt{-g} 
\left( g^{tt} 
 \bar{\phi}\phi \right) 
= (\sqrt{-g} g^{rr} \bar{\phi} \partial_r \phi)_{r=r_h+h} \,. \quad
\eea
The right hand side surface term vanishes  by $C_{r_h + h}=0$ with Dirichlet boundary condition, 
or Neumann boundary condition, but not for Dirichlet boundary condition with $C_{r_h + h} \neq 0$ nor 
ingoing/outgoing boundary condition. 

Remember that we need thermal equilibrium for the probe fields with the background black branes. 
This forces us that we cannot have a net energy flow between probe fields to the black branes therefore, 
only $C_{r_h + h}=0$ with Dirichlet boundary condition, 
or Neumann boundary condition are the possible choices.

Note that in the original derivation of the probe field free energy in \S II, as far as 
$g(E)$ in eq.~(\ref{gEdefinition}) is not qualitatively modified, 
we obtain the divergent free energy contributions. For $h \neq 0$,  
generic Dirichlet and Neumann boundary conditions are qualitatively the same in the sense that 
in both cases we obtain qualitatively the same $g(E)$. 
However ingoing/outgoing boundary conditions drastically change $g(E)$. 
Thus, the free energy with ingoing/outgoing or Dirichlet/Neumann boundary condition are completely 
different.  
Physically the spectrum obtained from the solutions with ingoing/outgoing boundary conditions 
has non-zero $\mbox{Im}(E) \neq 0$, so it is rather decaying object,   
due to the existence of the surface contribution.

Now we will consider the $h \rightarrow 0$ limit. 
For $h \to 0$, the solutions with the ingoing boundary condition 
are called the quasi-normal modes, \cite{Horowitz:1999jd, Iizuka:2003ad}. 
The number of low energy quasi-normal 
modes are finite \cite{Horowitz:1999jd} and $E$  have non-zero imaginary part. 
On the other hand, 
for the modes with the Dirichlet boundary condition with $h \to 0$, 
we can not take $h \to 0$ limit 
as a well defined limit. 
To see this, let us study the near the horizon geometry, $r \sim r_h$, 
$ds^2 =  - T (r- r_h) dt^2 + dr^2/T (r- r_h)$, By setting $\tilde r = dr/\sqrt{T(r- r_h)} 
=2  \sqrt{(r -r_h)/T}$, and $\theta = \pm i T t/2$, 
the metric approximates as 
\bea
ds^2 \sim d\tilde{r}^2 + \tilde{r}^2 d \theta^2 + d \vec{x}^2
\eea
which is a cigar geometry in Euclid signature.  
With the ansatz 
\bea
\phi = \phi(r) e^{i l\theta} \,,
\eea
Laplacian gives 
\bea
&& \Box \phi 
= \frac{1}{\tilde{r}} \partial_{\tilde{r}} (  {\tilde{r}} \partial_{\tilde{r}} \phi ) - \frac{l^2}{\tilde{r}^2}  \phi = 0
\eea
solution allows power law behavior 
$\phi(\tilde{r}) \sim {\tilde{r}}^l$,  $\phi \sim {\tilde{r}}^{-l}$. 
By setting $l = \mp 2 i E/T$, $\theta = \pm i T t/2$, we have 
\bea
\label{anlyticcontEuclidLorentz}
\phi \sim \tilde{r}^{ \mp 2 i E/T} e^{i E t} 
\sim (r -r_h)^{ \mp  i E/T} e^{i E t} \,.
\eea 
So 
the solutions are
\begin{eqnarray}
\phi &=&(r -r_h)^{-i {E}/{T}} h_- (1+{\cal{O}} (r-r_h)) \CR
&& + \, (r -r_h)^{i {E}/{T}} h_+ (1+{\cal{O}} (r -r_h)),
\end{eqnarray}
where terms proportional to $h_-$ and $h_+$ represent
the ingoing and outgoing modes, respectively.
Imposing the Dirichlet boundary condition, we have
\begin{eqnarray}
 \frac{h_+}{h_-}  
&=&-e^{-2 i {E}/T \ln h} \, (1+{\cal{O}} (h)) \,,  
\end{eqnarray}
in which the phase oscillate infinitely. 
Thus in addition to the fact that the number of low energy modes diverge in $h \to 0$ limit,  
we can not make the Dirichlet boundary condition well-defined in $h \to 0$ limit.

Let us discuss shortly the connection to the Euclidian case, since in that case, we have 
direct connection to the GKPW prescription \cite{Gubser:1998bc, Witten:1998qj} 
of the holographic Green function evaluation. 
For the $h\neq0$ case, we have two solutions. So we can take any boundary conditions by 
combining these two solutions. 
On the other hand, at $h = 0$, one solution diverges so we can {\it not} make any 
linear combination, in fact, it allows only {\it one} regular solution. 
This regular solution in Euclid signature goes to ingoing one in the Lorentzian signature as 
we have seen above eq.~(\ref{anlyticcontEuclidLorentz}). 
Actually by using this point and also the relationship between the 
Euclid Green function and retarded Green function 
\bea
G_R(\omega, \vec{k}) = G_E(\omega_E, \vec{k})|_{\omega_E = - i (\omega+ i \epsilon)}, 
\eea
it was shown in \cite{Iqbal:2009fd} that one can obtain 
the retarded Green function, which was first proposed in \cite{Son:2002sd} by bottom up approach, from 
the Euclid Green function using the GKPW prescription. 
For $h=0$ case, since it is admitted only one solution from the regularity 
in Euclid signature, we do not have much freedom to obtain other Green function by using GKPW prescription. 
In that sense, $h\neq0$ makes the situation better 
since it allows any linear combinations, as a results, any Green's functions.

\subsection{Dual field theory interpretation of the divergence}

Let us consider 
the background geometries dual to the confinement phase where the geometry has smooth IR ``cut-off''\footnote{This ``cut-off'' is 
the bulk IR endpoint of the geometry related to the confinement scale in the boundary theory, not the 
IR cut-off of the brick wall at the deconfinement phase.} at $r = r_c$ without 
black hole horizon. Examples for such geometries are  
 global AdS which has a spherical boundary, and for $R^4$ boundary, we have 
Klebanov-Strassler \cite{Klebanov:2000hb} or D4-brane geometry compactified 
on $S^1$ \cite{Witten:1998zw}.   
These geometries are dual to the confinement phase in boundary theories and 
related to the scale $r=r_c$, they admit nonzero confinement/deconfinement transition 
at finite $T = T_c$.  At temperature $T > T_c$, these geometries admit black hole horizon as 
\cite{Gubser:2001ri}.   
If we add probe branes on these setting \cite{Karch:2002sh}, 
then, in dual field theory, quark fields are introduced 
and we can study various meson confinement/deconfinement dynamics, see for example, 
\cite{Kruczenski:2003uq,Babington:2003vm,Sakai:2004cn,Aharony:2006da} and review \cite{Erdmenger:2007cm}. 
The probe fields are quarks and whose degrees of freedom are ${\cal O}(N)$ in the
Lagrangian level.\footnote{We neglect the effect of the massive open string modes 
by the non 't Hooft limit which we take, $l_s \sim l_p \ll 1$. 
These massive open strings are dual to 
massive vector mesons such as $\rho$ mesons.}
After a long time, the probe fields in the boundary theory will be 
thermalized with the gluons.\footnote{The backreaction of the probe
fields to the gluons will be small in the large $N$ limit.}
In the bulk theory, this is interpreted as 
a thermalization of the probe fields in the bulk with a black
hole.\footnote{
The backreactions to the black hole geometry by the probe fields 
will be small becuase the black hole mass is large.}
As we discussed in \S \ref{probebranemodelsubsection}, 
we consider the brick wall model for the probe scalar fields on the probe branes under such thermal equilibrium. 

For $T< T_c$, they are confined into the mesons.
The number of {\it low energy} spectrum of the mesons is finite, namely 
there is mass gap and the spectrum is discrete and low energy free energy does not diverge.  
This is consistent with the spectrum of the scalar
in the bulk Lorentzian (global) $AdS_d$ analysis.

For $T> T_c$, the quarks will be deconfined.
Therefore, in the gravity dual 
we expect that there are ${\cal O}(N)$ contribution,
which becomes divergent free energy 
in the large $N$ limit. This divergence is exactly what we have seen in the bulk analysis 
\S II and III in the $h\to 0$ limit, namely without the brick wall cut-off. 
Note that if we take the ingoing boundary condition, then we do not have such behavior
at least. 
Since it diverges, in the large but finite $N$ case, 
we need to put regulator near the horizon $r = r_h + h$, with $h\neq 0$.   
For a boundary condition there, we can take Dirichlet or Neumann boundary condition for example. 
This helps, since as we have seen in \S \ref{originofdiv}, 
the divergence comes from the divergence of the number of 
the {\it low energy} degree of freedom, namely $E\to 0$ modes. 
This $E$ corresponds to the 
energy conjugate to the boundary time, due to our choice of the 
ansatz (\ref{phiansatzveck})\footnote{The continuity of the low energy spectrum for the free energy 
divergence can be seen in other places.  
For example, by considering the scalar fields in 
Poincare coordinate pure AdS space, 
where we have zero temperature horizon, we can see that the spectrum becomes  
continuous and as a result, their free energy diverge, see for example, \cite{Erdmenger:2007cm}.}.   
Since the Dirichlet boundary condition suppresses the low energy mode divergence and make 
the spectrum from continuous to discrete, 
in field theory dual, this corresponds to introduction of the {\it IR cut-off} by the finite $N$ effect.  
This is the brick wall in a holographic context. 
Note that we expect that at infinite $N$ with nonzero coupling $\lambda$, 
the spectrum of the gauge theories at the deconfinement phase are continuous 
as is seen in the analysis \cite{Festuccia:2005pi,Festuccia:2006sa,Iizuka:2008hg,Iizuka:2008eb}. 
On the other hand, finite $N$ makes the spectrums discrete even at the deconfinement phase. 
The essential role of finite $N$ is to introduce the IR cut-off 
 and remove the divergent IR degeneracy by making the continuous spectrum into discrete spectrum 
 at the deconfinement phase,  
 and as a result, the free energy becomes finite.  
The brick wall with nonzero $h$ plays the role of the finite $N$ effect in the dual field theory, since 
finiteness of $N$ 
makes the spectrum discrete in the boundary theory at the deconfinement phase, and 
finiteness of $1/h$ 
makes the spectrum discrete in the bulk theory under the presence of a black hole.

The free energy will increase and diverge if we take $h \to 0$. 
However we know that the probe D7-branes introduce at most $O(N)$ (quarks) contributions in the 
Lagrangian level, and it cannot exceed the $O(N^2)$ contribution of the background D3 degrees of 
freedom.  
This gives the lower bound for $h$,  
where at least, 
naive picture allowed by local quantum field theory in curved space-time is not trustworthy. 
As we discussed in the end of \S III.C, 
this implies 
that the trust in local field theory for counting bulk d.o.f. under the given metric (\ref{thefirstmetricansatz}) will be lost at $h \sim l_p$, 
i.e. $r=r_h+{\cal O} (l_p)$
at least, for the black hole in the AdS space-time. 
More precisely, the semi-classical free field approximation for counting d.o.f. 
on the geometry will not be valid. 

In more generic setting, 
\bea
\label{NFphiFBB}
N^\delta F_\phi \sim F_{BB}  \sim N^\alpha \,, 
\eea 
we have seen that the IR cut-off $h$ is determined as 
\bea
\label{lphNrelation2}
h \sim (l_p)^{\frac{2(\alpha-\delta)}{\alpha}} 
\eea
in eq.~(\ref{lphNrelation}). 
Without $h$, the free energy of the probe field $F_{\phi}$ diverges. 
Since we expect that the metric and semi-classical description for counting d.o.f. are 
trustworthy from far infinity up to $O(1)$\footnote{$O(1)$ implies AdS scale or any other scales, which are 
independent on Planck or string scale.} close to the horizon 
$r \gtrsim r_h + O(1) \gg r_h + h$, 
this implies that 
$h$ must be at most positive power of Planck scale, especially, it cannot be independent on 
the Planck scale\footnote{Note that we are considering the limit where $g_s$ is fixed finite as $g_s \sim 1$ and 
$N \to \infty$ so that string scale and Planck scale are the same order and both are very small, $l_s \sim l_p \to 0$, compared with AdS curvature scale or large black hole 
curvature scale. Therefore in our limit, there is no distinction between Planck corrections and stringy corrections since both appear simultaneously in 
our limit.}.   
Therefore, it is reasonable to expect $\alpha > \delta$, so 
\bea
\label{alphadeltaargument}
F_{\phi} \sim N^{\alpha - \delta} \to \infty \quad \mbox{(in the large $N$ limit)}
\eea 
Therefore, $F_{\phi}$ always diverges by the positive powers of $N$. 

Of course, these arguments have several loopholes. One should note that  
we have {\it assumed} that at finite $l_p$, 
quantum gravity effects are not important at $r \gtrsim r_h + O(1) \gg r_h + h$, therefore we have squeezed the 
quantum gravity effects at the region $r \lesssim r_h + h$.  However we do not have good argument to 
justify this assumption. 
It is interesting to study this more concretely in holographic setting. 

Note that for the argument (\ref{alphadeltaargument}), we have assumed that 
the boundary theory does not have momentum space cut-off satisfying (\ref{smallLambdacut-off}). 
For D3-D7 \cite{Karch:2002sh} or D4-D8 \cite{Sakai:2004cn} system, the theory in UV approach the 
scale invariant theories and they do not admit any cut-off $\Lambda$. 
However if there is momentum space cut-off satisfying (\ref{smallLambdacut-off}), then the conclusion (\ref{alphadeltaargument}) 
break down and the cut-off $h$ could be bigger. 

Note also that 
the brick wall model requires the thermal equilibrium between the 
black hole and the probe field, 
and this is possible with the large black hole in the AdS space-time, 
since it allows positive specific heat.  
Moreover, the thermal equilibrium implies that 
there is no net energy flow between the probe fields and the black holes. 
In dual field theory, this is between probe quarks and the background gluons. 
Zero value Dirichlet boundary condition and Neumann boundary condition makes the surface term vanish, but 
ingoing and outgoing boundary conditions allow the nonzero surface term and therefore 
there are energy flows, so they will not allow a thermal equilibrium. 

The divergence of the free energy $F_{\phi}$ we have evaluated are, in fact, one-loop effects and 
the reader might wonder at the tree level contribution. 
In fact, as is seen in \cite{Mateos:2006nu, Mateos:2007vn} 
the classical Dirac-Born-Infeld (DBI) action evaluated on the Euclidean black hole background yields the $O(\lambda N_f N)$ results. 
Note that this is the same as Gibbons-Hawking prescription \cite{Gibbons:1976ue}, and 
$O(\lambda N_f N)$ contribution is due to the horizon existence. 
This is correct in the $N\to \infty$ limit with $\lambda$ fixed. 
On the other hand, in the brick wall context where $N$ is large but rather finite, 
we do not expect the sharp notion of the horizon existence; full quantum gravity effects 
become important and geometry is not trustworthy slightly outside of the horizon, which is described by the 
brick wall at $r= r_h + h$. 
So tree level contributions are actually expected to be $O(1)$
\cite{Mukohyama:1998rf} in our case. 
Therefore this one-loop divergence can be dominant\footnote{%
Usually, the 1-loop effects give  $O(1)$ contributions. 
However, in our case the number of the low enegy modes 
are infinite and $O(1) \times \infty$ contribution 
in the $N \rightarrow \infty$ limit, i.e. $h \rightarrow 0$ limit. 
Then, because $N$ is finite and $h$ is finite, 
the 1-loop effects can give $O(N)$.
On the other hand, in the usual $1/N$ expansion picture, 
tree level contribution is $O(N)$.}. 
We emphasize that 
as we have interpret, this IR divergence is physical 
in the sense that it is associated with the continuity of the spectrum in the boundary theory. 
We will discuss more in the discussion section for the finite $N$ effect as an IR cut-off. 

\subsection{Difficulty of having a fundamental ``electron'' in the dual field theories}

We have seen that under the black hole background, there are universal nature for the 
the probe free energy $F_\phi$; it always diverges 
due to the continuous gapless spectrum. 
The universality of the brick wall implies that all the probe fields show 
divergent free energy by the same mechanism. What does this mean in a boundary context? 
We interpret this 
as a signal that all the probe fields which are singlet at the confinement phase, show the 
continuous gapless spectrum 
with large IR degeneracy 
at the deconfinement phase. So they are actually not the fundamental fields such as electrons but 
rather they are at most composite fields. 

What this brick wall model argument is suggesting is that {the holographic dual field theory 
can have gauge-singlet fields ($SU(N)$ singlet) only as composite fields, 
like mesons\footnote{This is the argument for theories which have some UV cut-off energy scale 
and we assume that this UV cut-off is   
much bigger than the confinement-deconfinement transition.}. 
In other words, the brick wall model suggests that the gauge singlet components should always  
show the confinement-deconfinement transition and their degrees of freedom (in the Lagrangian level) should scale by a positive power of $N$, 
therefore, we can have electron like objects as composite objects like mesons, or fermionic superpartners of mesons to make them fermionic. 
 
To see this, let's consider adding a $SU(N)$ gauge-singlet fundamental field (in the Lagrangian level), 
like an electron field, to the $SU(N)$ QCD like theory. 
Suppose that the gauge-singlet fundamental field becomes thermal equilibrium, then we can consider 
its free energy. 
The contribution to the free energy from the gauge singlet field should be independent on $N$, 
so it is always $O(1)$ even after the $SU(N)$ QCD like theory show 
the confinement-deconfinement transition.  
However this contradicts the gravity analysis; 
if we consider the free energy of the bulk scalar field, which is dual to the gauge singlet field,   
it shows the divergence without the brick wall cut-off due to the near horizon throats 
in the vicinity of the black holes with $h = 0$. 
With IR cut-off $h \neq 0$, free energy behaves 
as $\sim 1/h$. 
Note that this divergence is due to the almost continuity of the spectrum in the IR, 
therefore its degrees of freedom cannot be $O(1)$. Therefore, they cannot be gauge singlet fields after the confinement-deconfinement transition, but rather they are composite fields 
at most\footnote{One possible loophole of this argument is that this discrepancy is due to the strong 
coupling $\lambda$ effect, since gravity is at large $\lambda$ but gauge theory is at small $\lambda$, 
so free energy behaves as a positive power of $\lambda$, instead of $N$.   
However in the limit we are considering, where $g_{YM} = O(1)$, we do not distinguish the large $\lambda$ effect and large $N$ effect since $\lambda \sim N$.}. 

The fact that they always show these divergence is due to the universality of the gravity. 
Gravity couples to all the fields, furthermore gravitational theory, which has asymptotically AdS boundary,  
always form a large black hole at high temperature.  
Therefore, as we have seen in \S II and \S III, any bulk fields dual to the gauge-singlet matter fields 
obey their equations of motion, and 
their free energy always 
show the divergence due to the localized modes near the black hole horizon, 
at least in large $N$ limit. 
 

How much the cut-off is necessary in the gravity side is evaluated from the field theory sides. 
If we add the glueballs, then their free energy and entropy scale as $O(N^2)$ at the 
deconfinement phase. This makes the 
free energy of the dual bulk scalar fields the same order as background black brane free 
energy (which is dual to the gluons), and this fix the $h$ to be Planck scale in the invariant distance. 
On the other hand,  if we introduce the probe brane as we discussed in previous subsection, then free energy of the meson scalar is $O(N)$ 
at the deconfinement phase, which is suppressed by $N$ from the black brane entropy. This yields the 
new IR cut-off scale. Again, note that the IR cut-off $h$ in eq.~(\ref{lphNrelation2}) is {\it not} invariant distance 
but rather distance in terms of $r$-coordinate.

\subsection{Path-integral measure from Euclidian analysis} 

So far we have seen the divergence from the Lorentzian signature. Since after the analytic continuation, 
the bulk geometry near the black branes or holes become simply smooth cigar geometry where 
no metric components shows coordinate singularities, unlike Loretzian signature. Therefore 
one might wonder in Euclid signature, these divergences disappear. However since we have 
seen the divergences are physical from dual field theory viewpoint, it should not disappear even 
after the analytic continuation.  
Here we will conduct the partition function evaluation by Euclidian path-integral method and 
see that the divergences still exist. This subsection is mainly a review. 

For the partition function evaluation in Euclidian path-integral analysis, 
we can consider the usual Matsubara formalism.
We will see that 
the path-integral measure for this is different from the diffeomorphism invariant one 
with the conformal transformation by 
$g_{tt}$ \cite{Barbon:1994ej, Giddings:1993vj, Emparan:1994qa, de Alwis:1995cr}. 

By interpreting partition function as Euclid time evolution by $\beta$, 
\bea
\label{Epartitionfn}
Z  
= \Tr e^{-\beta H}  = \int [d \phi] e^{- \int d \tau L_E (\phi)}
\label{Euclidpartitionfunction}
\eea
We will evaluate this path-integral in this subsection. 

For that purpose, let us recall the analysis in Lorentzian signature. 
We will define the complete set motivated by the Lorentzian signature Laplacian $\nabla^2_L$,  
\bea
\nabla_{L}^2   =  g^{tt} \partial_t^2 + \nabla_{r,x}^2 = g^{tt} \left( \partial_t^2 + g_{tt} \nabla_{r,x}^2 \right) 
\eea
such that complete sets we define are 
\bea
\label{completesets}
g_{tt}(r) \nabla_{r,x}^2 \phi_{n,r,x} &=& E_n^2  \phi_{n,r,x} \,,
\eea
with the normalization 
\bea
- \int  dr dx \sqrt{-g}g^{tt}  \phi_{n,r,x}  \phi_{m,r,x}  &=&  \delta_{n,m} \,.
\eea
Note that this gives precisely the eq.~(\ref{scalarwaveeq1}).

Then we expand the field as 
\bea
\phi(t,r,x) = \sum_n c_n(t)   \phi_{n,r,x} \,. 
\eea
The solutions of the equation of motion are given by 
$c_n(t)=c_n e^{i E_n t}$.
Then the action is
\begin{eqnarray}
 \int dt L= \int dt \sum_n \left( 
\dot{c}_n(t)^2 - E_n^2 c_n(t)^2 
\right),
\end{eqnarray}
is the harmonic oscillators for quantum mechanical modes $c_n(t)$.
Thus, the Hamiltonian is 
\bea
H = \sum_{n=0}^{\infty}  
(m_n+\frac{1}{2}) E_n
\eea
for the eigenstates, 
\bea
|\{m_n\}>  = \Pi_{n=0}^{\infty} (a_n^\dagger)^{m_n} |0>  \,.
\eea
labeled by $\{m_n \}$ where $m_n=0,1,2,\cdots$,
and this reduces to the \S II analysis.

Motivated by this Lorentzian analysis, we will evaluate the eq.~(\ref{Euclidpartitionfunction}).  
In the path-integral conducted in Euclid signature, 
we will expand the field as
\bea
\phi(\tau,r,x) = \sum_n c_n(\tau)   \phi_{n,r,x} \,,
\eea
where we choose 
the complete sets $\phi_{n,r,x}$  similar to the one defined in eq.~(\ref{completesets}) in Lorentzian case, {i.e.,} we choose, 
\bea
\label{Euclidcompletesets}
- g_{\tau\tau}(r) \nabla_{r,x}^2 \phi_{n,r,x} = E_n^2  \phi_{n,r,x} \,.
\eea

The Euclid Lagrangian $L_E$ gives 
\bea
\int d\tau L_E(\phi) &=&  \int d\tau dr d\vec{x} \sqrt{g}g^{\tau \tau}  \sum_{m,n} \left(  \dot{c}_n(\tau) \dot{c}_m(\tau)  \phi_{n,r,x}  \phi_{m,r,x}  \right. \CR
&&  \left.
+ (E_m)^2 c_n(\tau)  c_m(\tau)   \phi_{n,r,x}  \phi_{m,r,x} \right) \nonumber \\
&=&  \int d\tau \sum_n \left(  \dot{c}_n(\tau)^2 + E_n^2 c_n(\tau)^2 \right) \nonumber \\
&=& \sum_n \sum_m \left( (\frac{2 \pi m}{\beta})^2 + E_n^2 \right) (c_{nm})^2 
\label{EuclidLevaluation1}
\eea
where we have used the fact that periodicity under the 
$\tau \to \tau +\beta$ from the trace property to expand
$c_n(\tau) \sim \sum_{m \in {\bold Z}} c_{nm} e^{\frac{2 \pi i}{\beta} m \tau}$ 
and the completeness condition
\bea
\label{nondiffcompletenorm}
 \int  dr dx \sqrt{g}g^{\tau \tau}  \phi_{n,r,x}  \phi_{m,r,x}  = \delta_{n,m}
\eea
as a normalization.  

This yields the path-integral evaluation of the partition function
eq.~(\ref{Epartitionfn}) up to a numerical normalization constant, 
\bea
Z = \Pi_n  \Pi_{m \in {\bold Z}} 
\left( (\frac{2 \pi m}{\beta})^2 + E_n^2 \right)^{-\frac{1}{2}} \,.
\eea

Consider the background which is pure AdS. Then the partition function determined by this 
prescription gives the thermodynamical partition function of the, for example, mesons.   
This justifies that prescription above gives the correct answer in AdS/CFT viewpoint. 

On the other hand, above prescription gives the divergent contribution on the path-integral method 
in view of Euclid prescription. This divergence is due to the 
\bea
\label{cigercenter}
g^{\tau \tau}  = \infty  \,,
\eea
for the complete sets (\ref{Euclidcompletesets}), since then, 
we have large $n$ mode with fixed energy and this yields the 
free energy divergence. The large $n$ mode has bigger 
value for the $r$ derivatives. The existence of huge $n$ with fixed $E \sim T$ is the origin 
of the divergence. 

Naively one might think that because in Euclid signature, the black hole is replaced by the 
cigar geometry so there is no warped factor, at least locally. 
However, since the normalization must be chosen in such a way (\ref{Euclidcompletesets}), 
the effect of (\ref{cigercenter}) still induces the divergence for the partition function. 

If we used the different complete sets, for example, 
\bea
\label{wrongcompletesets}
- \nabla_{r,x}^2 \phi_{n,r,x} = E_n^2  \phi_{n,r,x}
\eea
this does not diagonalize the path-integral, therefore it does not reduce to the simple determinant, 
unlike (\ref{EuclidLevaluation1}).

On the geometry far outside the horizon, we believe usual quantum field 
theory in curved space-time should holds as a good approximation. 
This means that we should path-integrate the scalar field
on the geometry as usual. However as we have seen, 
the treatment for the basis choice in eq.~(\ref{Euclidcompletesets}) and (\ref{nondiffcompletenorm}) treats diffeomorphism non-invariant way, 
since we treat $g^{\tau\tau}$ special and this induces divergence as eq.~(\ref{cigercenter}). 
The reason for treating $g^{\tau\tau}$ special must be clear. In order to define energy or temperature for thermal partition functions, 
we have to specify their conjugate, which is time, that is why we treat metric component $g^{\tau\tau}$ special.

\section{Discussion}

Probe fields in the bulk, which are dual to the 
probe fields like meson fields, show the divergent free energy due to the near horizon throat 
of black holes. 
The origin of the free energy divergence is due to the 
large degeneracy of the modes, living near the throat region of the horizon $r - r_h \gtrsim h$, with 
\bea
\label{bigguys}
E \lesssim \sqrt{\frac{T}{h}} \,, \quad \vec{k}_{local}^2 \lesssim \frac{T}{h} \,.
\eea
in bulk theory viewpoint. We have seen that 
the value of cut-off $h$ depends on the microscopic theories.  
In the microscopic system where the degrees of freedom for background black brane $F_{BB}$ and 
probe fields $F_{\phi}$ is given by 
\bea
\label{DpDp'NalphaNdelta}
F_{BB} \sim N^{\alpha} \,, \quad F_{\phi} \sim N^{\alpha - \delta} \,, 
\eea
then the appropriate IR cut-off $h$ is 
\bea
h \sim (l_{p})^{\frac{2 (\alpha - \delta)}{\alpha}}
\eea
otherwise, semi-classical approximation over-count the degrees of freedom 
as we have seen in \S IV-B. 
For example, in the D$p$-D$p'$ brane setting with $(p,p')=(3,7)$, where we have $N_c$ D$p$ branes with probe $N_f$ D$p'$ branes, we have $\alpha = 2$ and $\delta = 1$. 
These probe D$p'$-branes introduce microscopic 
quark fields which have no mass gap and continuous spectrum in the boundary 
theory at the deconfinement phase. 
In this way, we have seen that these divergences are not pathological, but  
rather physical consequence, at least for the probe brane models from the holographic viewpoint. 
Note that in the bulk viewpoint, if we do not remove the modes 
\bea
\label{diverging_guys}
E \gg \sqrt{\frac{T}{h}} \,, \quad \vec{k}_{local}^2 \gg \frac{T}{h} \,,
\eea
then the free energy of the probe field becomes bigger than the microscopic counting (\ref{DpDp'NalphaNdelta}), 
which is inconsistent. Taking only the modes satisfying (\ref{bigguys}) but not the modes 
satisfying (\ref{diverging_guys}), we obtain a large but non diverging free energy as (\ref{DpDp'NalphaNdelta}).

In \S \ref{originofdiv} we have seen that 
after the $\vec{k}$ integration, we have enhancement for $1/h$ by the 
contributes $g_{rr} \sim 1/h$ factor. 
This implies that large $|\vec{k}|^2 \sim {{T}/{h}}$ does not give large $E$ due to the near horizon warping effect.  
In dual field theory, this is interpreted as due to the interaction with the background gluon plasmas, 
which are dual to black holes. 

In the dual field theory, the semi-classical limit  
corresponds to the large $N$ limit and then 
the free energy for the deconfinement phase quarks 
should scale as ${\cal O}(N)$, which is divergent. From this microscopic analysis, 
it should be answered what are the cut-off effects 
in the gravity side, which corresponds to the finite $N$ effects. 
We have seen that once we specify 
the microscopic degrees of freedom 
for both probe fields and background black branes, then the cut-off is obtained as 
appropriate positive power of the Planck scale. 

What we have seen is that the semi-classical path-integral method for counting 
the degrees of freedom diverges in the large $N$ limit from both bulk and boundary viewpoint. 
But on large but finite $N$, bulk semi-classical path-integral method over-counts 
due to the modes eq.~(\ref{diverging_guys}). 
Therefore, this implies that the semi-classical approximation 
and our trust for the naive picture of the black hole geometries might break down  
for $r-r_H \lesssim h$.  
The point that the 
semi-classical approximation breaks down near the horizon is indeed what we expect from
the recent firewall argument or fuzzball argument, 
\cite{Mathur:2005zp, Mathur:2009hf, Mathur:2011uj, Almheiri:2012rt},  
although, these argument are 
based on other viewpoints: the inconsistency between unitarity of black hole evaporations and local quantum field theory 
in curved space near the horizon. 
Our argument is different; it is rather based on the over-counting issues 
of the degrees of freedom 
in the semi-classical limit, 
and also on 
how the finite $N$ or $l_{p}$ effects should appear in the bulk.

In full quantum gravity, the geometric picture probably breaks down.  
However, it might be reasonable to {\it assume} that at least 
the geometry is very well approximated 
to the black hole geometry as far as $r \gg r_H$. 
Therefore, it might be 
reasonable to expect the probe field free energy divergence is cured and  
the modes (\ref{diverging_guys}) are effectively removed  
by the quantum gravity effects at $r \lesssim r_H + h$.  

The role of the finite $N$ effect in the boundary theory reduces the degrees of freedom, 
compared with $N \to \infty$ limit. 
From a boundary theory viewpoint, 
these finite $N$ regulator effectively reduce the $\infty$ by $\infty$ matrices  
into finite $N$ by $N$.     

Note that we expect that at infinite $N$ with nonzero coupling $\lambda$, 
the spectrum of the gauge theories at the deconfinement phase are continuous 
as is seen in the analysis \cite{Festuccia:2005pi,Festuccia:2006sa,Iizuka:2008hg,Iizuka:2008eb}. 
On the other hand, finite $N$ makes the spectrums discrete even at the deconfinement phase. 
Therefore, the reduction of degrees of freedom by finite $N$ makes the spectrum from continuous to discrete 
even at the deconfinement phase, and this discreteness is crucial to see the Poincare recurrence of the 
time-evolution of the unitarity.  Apparent non-unitarity of the black hole information loss is due to the 
continuity of the spectrum at the deconfinement phase.  
With infinite $N$ the phase space volume is expected to be infinite\footnote{Note that if the spectrum is discrete even at $N \to \infty$ limit as is $\lambda \to 0$ case, then the phase space volume becomes finite and the time-scale of the Poincare recurrence is finite, contradicted with the classical information loss picture.} while with finite $N$, the phase space volume is expected to be finite, and no physical quantities should not diverge. 
If the spectrum is continuous, the 
Green functions do not show Poincare recurrence and decay in late time as \cite{Festuccia:2005pi,Festuccia:2006sa,Iizuka:2008hg,Iizuka:2008eb}. 
In that aspects,  our argument is also implicitly relying on the unitarity nature of the 
quantum mechanical system.  


From a bulk theory, both introducing a non-zero cut-off $h$ 
and a Pauli-Villars regulator like 
\cite{Demers:1995dq} have an effect to reduce the UV degrees of freedom
above Planck scale (measured in the local Lorentz frame)
as eq.~(\ref{diverging_guys}) 
with eq.~(\ref{hTMsqrelation}). Therefore, by introducing the brick walls, the spectrum  also change 
from the continuous one to discrete one.  
The point that these effects are UV in bulk theory but are IR in boundary theory is, due to the UV/IR mixing in holography \cite{Susskind:1998dq,Peet:1998wn}.  
Therefore, finite $N$ effects are IR cut-off in a boundary theory to make the spectrum discrete, and in the bulk, this is what 
brick wall does with $h \neq 0$. Brick wall effects with finite $1/h$ are the same as finite $N$ effects in the 
boundary theory and 
these are IR cut-off effects for the boundary theory spectrum.

The reductions of degrees of freedom from classical limit due to the non-perturbative quantum gravity effects 
are expected to be generic. 
In fact, 
they are seen using the gauge/gravity correspondence for many situations. 
Finite $N$ matrix model for $M$-theory \cite{Banks:1996vh,Susskind:1997cw,Maldacena:1997re,Polchinski:1999br} yields the quantum gravity reduction.  
In \cite{Lin:2004nb}, 
the reduction by finite $N$ occurs due to the 
``droplet'' quantization, also in \cite{Iizuka:2008eb}, the discreteness nature of the Young tableaux reduces the degrees of freedom. 
Such reductions of degrees of freedom in full quantum gravity 
is important and the cut-off $h$ of the brick wall model reflects this reduction. 
This suggestion might be important 
since this give another estimate 
of the where the locality breaks down, and deviation from classical gravity limit occurs.

Originally, the brick wall 
was introduced for regularization of 
the divergence of the free energy. 
This divergence can be absorbed into 
the renormalized Newton constant \cite{Susskind:1993if,Demers:1995dq}.
If our goal is simply to obtain finite numbers for the free energy of the probe fields, 
the classical DBI action evaluation near the horizon 
gives the correct non-diverging free energy in the large $N$ leading order \cite{Mateos:2006nu, Mateos:2007vn}, and 
the brick wall may not be needed. 
From holographic boundary theory viewpoint, 
this Newton constant renormalization can be 
regarded as an effective $N$ change as 
\bea
F_{BB}(N^2) + F_{\phi}(N) = F_{BB}(N^2_{eff}) \,, 
\eea 
so that $O(N)$ contribution of $F_{\phi}$ is absorbed as an effective $N$ change into $N_{eff}$ for $F_{BB}$. 
This might be meaningful  
if $N$ is infinite and one can be regarded $1/N$ as an almost continuous parameter. 
However in finite $N$, it is unclear if such an effective $N$ change makes sense. 
Similarly, the problem behind the divergence
is the {\it almost} continuous low energy spectrum in large $N$. 
Thus, 
the real advantage of the  
brick wall picture is for a finite $N$, i.e. a finite $l_p$ theory, where
the naive classical picture of horizons do not work. 
It is interesting to investigate more on 
the difference between the large $N$ picture and 
the finite $N$ picture, where non-perturbative quantum gravity effects are dominating.

\acknowledgments
We would like to thank especially 
Dan Kabat, Shinji Mukohyama, and Rob Myers for helpful discussions and comments on the 
draft. 
We would also 
like to thank Masanori Hanada, Kinya Oda, Naoki Sasakura, Masaki Shigemori, Tadashi Takayanagi, Tomonori Ugajin, and 
Satoshi Yamaguchi for helpful discussions. 
N.I. was supported in part by JSPS KAKENHI Grant Number 25800143. 
S.T. was supported in part by JSPS KAKENHI Grant Number 23740189.



\end{document}